\documentstyle[12pt]{article}
\newcommand{\be}{\begin{equation}}
\newcommand{\ee}{\end{equation}}
\begin{document}
\def\theequation{\arabic{section}.\arabic{equation}}
\begin{titlepage}
\title{Conformal transformations in classical gravitational theories and
in cosmology}
\author{Valerio Faraoni$^{1}$, Edgard Gunzig$^{1,2}$ and Pasquale 
Nardone$^1$\\ \\
{\small \it $^1$ RGGR, Facult\'{e} des Sciences} \\
{\small \it Campus Plaine, Universit\'{e} Libre de Bruxelles}\\
{\small \it Boulevard du Triomphe, CP 231}\\
{\small \it 1050 Bruxelles, Belgium}\\
{\small \it $^2$ Instituts Internationaux de Chimie et de Physique Solvay}
}
\date{}
\maketitle   \thispagestyle{empty} 
\begin{abstract}

In recent years, the use of conformal transformation techniques has become
widespread in the literature on
gravitational theories alternative to general relativity, on
cosmology, and on nonminimally coupled scalar fields.
Tipically, the transformation to the Einstein frame is
generated by a fundamental scalar field already present in the theory.
In this context, the problem of which conformal
frame is the physical one has to be dealt with and, in the general case, it
has been clarified only recently; the formulation
of  a theory in the ``new'' conformal frame leads to departures from
canonical Einstein gravity. In this article, we review the literature on
conformal transformations in classical gravitational theories and
in cosmology, seen both as purely mathematical tools and as maps with
physically relevant aspects. It appears particularly urgent to refer the
analysis of experimental tests of Brans--Dicke and scalar--tensor
theories of gravity, as well as the predictions of
cosmological inflationary scenarios, to the physical conformal
frame, in order to have a meaningful comparison with the
observations.
\end{abstract}
\vspace*{0.3truecm}
\begin{center}
IUCAA 24/98
\end{center}
\begin{center} To appear in {\em Fundamentals of Cosmic Physics}.
\end{center}
\end{titlepage}   \clearpage
\begin{center} {\bf Contents} \end{center}
\begin{itemize}
\item Notations and conventions
\item 1. Introduction
\item 2. Conformal transformations as a mathematical tool
\item 3. Is the Einstein frame physical~?
\item 4. Conformal transformations in classical theories of gravity
\item 5. Conformal transformations in cosmology
\item 6. Experimental consequences of the Einstein frame reformulation of
      gravitational theories
\item 7. Nonminimal coupling of the scalar field
\item 8. Conclusions
\item Acknowledgments
\item References
\end{itemize}
\clearpage

\section*{Notations and conventions}

The notations and conventions used in this paper are as
follows: the metric signature is --~+~+~...~+. To facilitate the
comparison with the literature on inflationary cosmology, we use units in
which the speed of light and the reduced Planck constant assume the value
unity. $G$ is Newton's constant and the Planck mass is $m_{pl}=G^{-1/2}$
in these units. Greek indices assume the values 0,~1,~2,~...,~$n-1$, where
$n$ is the dimension of spacetime. When $n=4$, small Latin indices assume
the values 1,~2,~3. While we allow for $n$ spacetime dimensions (only one
of which is timelike), in most of this paper the value $n=4$ is assumed,
except when discussing Kaluza--Klein and string theories prior to
compactification.

A comma denotes ordinary differentiation, and $\nabla_{\mu}$ is the
covariant derivative operator. Round and square brackets around indices
denote, respectively, symmetrization and antisymmetrization, which
include division by the number of permutations of the indices: e.g.
$A_{( \mu\nu )}=\left( A_{\mu\nu}+A_{\nu\mu} \right) /2$.
The Riemann and Ricci tensors are given in terms of the Christoffel
symbols $\Gamma_{\alpha\beta}^{\delta}$ by
$$
{R_{\alpha\beta\gamma}}^{\delta}=\Gamma_{\alpha\gamma
,\beta}^{\delta}-\Gamma_{\beta\gamma , \alpha}^{\delta}
+\Gamma_{\alpha\gamma}^{\sigma} \Gamma_{\sigma\beta}^{\delta}-
\Gamma_{\beta\gamma}^{\sigma}\Gamma_{\sigma\alpha}^{\delta} \; ,
$$
$$R_{\mu\rho}=
\Gamma^{\nu}_{\mu\rho ,\nu}-\Gamma^{\nu}_{\nu\rho ,\mu}+
\Gamma^{\alpha}_{\mu\rho}\Gamma^{\nu}_{\alpha\nu}-
\Gamma^{\alpha}_{\nu\rho}\Gamma^{\nu}_{\alpha\mu} \; , $$
and $R\equiv g^{\alpha\beta}R_{\alpha\beta}$ is the Ricci curvature.
$\Box \equiv g^{\mu\nu}\nabla_{\mu}\nabla_{\nu}$ is d'Alembert's
operator. A tilde denotes quantities defined in the Einstein 
frame, while a caret denotes quantities defined in a higher--dimensional
space prior to
the compactification of the extra dimensions.

\section{Introduction}

If $(M, g_{\mu\nu})$ is a spacetime, the point--dependent
rescaling of the metric tensor
\be                       \label{CT}
g_{\mu\nu} \rightarrow \tilde{g}_{\mu\nu}=\Omega^2 g_{\mu\nu} \; ,
\ee
where $\Omega =\Omega(x) $ is a nonvanishing,
regular function, is called a {\em
Weyl} or {\em conformal transformation}. It
affects the lengths of time [space]--like
intervals and the norm of time [space]--like vectors, but it leaves
the light cones unchanged: the spacetimes $(M, g_{\mu\nu})$ and
$(M, \tilde{g}_{\mu\nu})$ have the same causal structure. The converse is
also true (Wald 1984). If $v^{\mu}$ is a null, timelike, or spacelike
vector with respect to the metric $g_{\mu\nu}$, it is
also a null, timelike, or spacelike vector, respectively, in the rescaled
metric $\tilde{g}_{\mu\nu}$.

Denoting by $g$ the determinant det$(g_{\mu\nu})$ one has, under the
action of (\ref{CT}),  $\tilde{g}^{\mu\nu}=\Omega^{-2}
g^{\mu\nu}$ and $\tilde{g} \equiv \mbox{det}\left(
\tilde{g}_{\mu\nu}\right) =\Omega^{2n} g $.
It will be useful to remember the transformation properties of the
Christoffel symbols, Riemann and Ricci tensor, and of the Ricci
curvature $R$ under the rescaling (\ref{CT})
(Synge 1955; Birrell and Davies 1982; Wald 1984):
\be
\tilde{\Gamma}^{\alpha}_{\beta\gamma}=
\Gamma^{\alpha}_{\beta\gamma}+\Omega^{-1}\left(
\delta^{\alpha}_{\beta} \nabla_{\gamma}\Omega +
\delta^{\alpha}_{\gamma} \nabla_{\beta} \Omega
-g_{\beta\gamma}\nabla^{\alpha} \Omega \right) \; ,
\ee
\begin{eqnarray}
& & \widetilde{ {R_{\alpha\beta\gamma}}^{\delta}}=
{R_{\alpha\beta\gamma}}^{\delta}+2\delta^{\delta}_{[\alpha} \nabla_{\beta
]}\nabla_{\gamma} ( \ln \Omega )
-2g^{\delta\sigma} g_{\gamma [ \alpha}\nabla_{\beta ]}\nabla_{\sigma}
( \ln
\Omega ) +2 \nabla_{[ \alpha} ( \ln \Omega ) \delta^{\delta}_{\beta ]}
\nabla_{\gamma}( \ln \Omega ) \nonumber \\
& & -2\nabla_{[ \alpha}( \ln \Omega )g_{\beta ]
\gamma} g^{\delta \sigma} \nabla_{\sigma}( \ln \Omega )
-2g_{\gamma [ \alpha} \delta^{\delta}_{\beta ]} g^{\sigma \rho}
\nabla_{\sigma} ( \ln \Omega ) \nabla_{\rho} ( \ln \Omega ) \; ,
\end{eqnarray}
\begin{eqnarray}
& & \tilde{R}_{\alpha\beta }=R_{\alpha\beta }
-(n-2) \nabla_{\alpha}\nabla_{\beta }
( \ln \Omega )
-g_{\alpha\beta } g^{\rho\sigma } \nabla_{\rho} \nabla_{\sigma}
( \ln \Omega )
+(n-2)  \nabla_{\alpha} ( \ln \Omega ) \nabla_{\beta}( \ln \Omega )
\nonumber \\
& & -(n-2) g_{\alpha\beta }\, g^{\rho\sigma}
\nabla_{\rho}( \ln \Omega )  \nabla_{\sigma}( \ln \Omega ) \; ,
\end{eqnarray}
\be
\tilde{R} \equiv \tilde{g}^{\alpha\beta} \tilde{R}_{\alpha\beta }=
\Omega^{-2} \left[ R-2 \left( n-1 \right) \Box \left( \ln \Omega \right) -
\left( n-1 \right) \left( n-2 \right)
\frac{g^{\alpha\beta} \nabla_{\alpha} \Omega \nabla_{\beta}
\Omega}{\Omega^2}
\right] \; ,
\ee
where $n$ ($n \geq 2 $) is the dimension of the spacetime manifold $M$.
In the case $n=4$, the scalar curvature has the expressions
\be
\tilde{R} =\Omega^{-2} \left[ R-\frac{6 \Box \Omega}{\Omega} \right] =
\Omega^{-2} \left[ R-\frac{12 \Box ( \sqrt{\Omega})}{\sqrt{\Omega}} -
\frac{3g^{\alpha\beta} \nabla_{\alpha} \Omega \nabla_{\beta} \Omega}
{\Omega^2}     \right]  \; ,
\ee
which are  useful in many applications. The Weyl tensor
${C_{\alpha\beta\gamma}}^{\delta} $ (beware of the position of the
indices~!) is conformally invariant:
\be
\widetilde{ {C_{\alpha\beta\gamma}}^{\delta}}=
{C_{\alpha\beta\gamma}}^{\delta} \; ,
\ee
and the null geodesics are also conformally invariant (Lorentz 1937).
The conservation equation $\nabla^{\nu} T_{\mu\nu} =0 $ for a symmetric
stress--energy tensor $T_{\mu\nu} $ is not conformally invariant
unless the
trace $T \equiv {T^{\mu}}_{\mu}$ vanishes (Wald 1984). The Klein--Gordon
equation
$\Box \phi =0 $ for a scalar field $\phi$ is not conformally invariant, but
its
generalization
\be
\Box \phi-\frac{n-2}{4(n-1)} \,R \, \phi=0
\ee
($n \geq 2$) is conformally invariant (note that the introduction of a
nonzero cosmological constant in the Einstein action for gravity creates
an effective mass, and a length scale, in the Klein--Gordon equation,
which spoils the conformal invariance (Madsen 1993)). Maxwell's
equations in four dimensions are
conformally invariant (Cunningham 1909; Bateman 1910), but the
equations for the electromagnetic four--potential are
not (it is to be noted that, at the quantum level, the conformal invariance
of the Maxwell equations  may be broken by quantum
corrections like the generation of mass or the conformal anomaly).
The conditions for conformal invariance of fields of
arbitrary spin in any
spacetime dimensions were discussed in (Iorio {\em et al.} 1997).

In this review paper, we will limit ourselves to consider special conformal
transformations, in which the dependence of the conformal factor $\Omega
(x) $ on the spacetime point $x$ is obtained via a functional dependence
(usually a power law) on a scalar field $\phi (x) $ present in the theory:
\be     \label{specialCT}
\Omega (x) =\Omega \left[ \phi (x) \right] \; .
\ee
A redefinition of the scalar field $\phi$ accompanies the conformal
transformation (\ref{CT}).
Theories in which a fundamental scalar field appears and generates
(\ref{CT})
include scalar--tensor and
nonlinear theories of gravity (in which $\phi$ is a Brans--Dicke--like
field) and
Kaluza--Klein theories (in which $\phi$ is the determinant of the metric
of the extra
compact dimensions). Fundamental scalar fields in quantum theories include
$SO(N)$ bosons in dual models, Nambu--Goldstone bosons, Higgs fields, and
dilatons in superstring theories. In addition, almost
all\footnote{The exception is $R^2$ inflation
(Starobinsky 1980; Starobinsky 1986; Maeda, Stein--Schabes
and Futamase 1989), in which the
Lagrangian term $R^2$ itself drives inflation. However, a scalar field is
sometimes added to this scenario to ``help'' inflation
(Maeda 1989; Maeda, Stein--Schabes and Futamase 1989) and the scenario is
often recast as power--law inflation by using a conformal transformation
(Liddle and Lyth 1993).} scenarios of cosmological
inflation (Linde 1990; Kolb and Turner 1990; Liddle and Lyth 1993; Liddle 
1996) are based on scalar fields, either in the context of a classical or
high energy theory, or in a phenomenological approach in which a scalar
field is introduced as a source of gravitation in the field equations of
the theory (usually
the Einstein equations of general relativity). By means of a transformation
of the form (\ref{CT}), many of these scenarios are
recast in the form of Einstein gravity with
the scalar field(s) as a source of gravity and a power--law inflationary
potential. The investigation of this
mathematical equivalence has far--reaching consequences, and in many cases
the
mathematical equivalence provides a means to go from a physically
inconsistent
theory to  a viable one. Unfortunately, the use of conformal transformations
in gravitational theories is haunted by confusion and ambiguities,
particularly in relation
to the problem of identifying the conformal frame which correctly
describes the physics.
Despite early work on the subject, confusion still persists in the
literature
and considerably detracts from papers that use conformal
techniques incorrectly.

It must be stressed that, in general, conformal transformations are not
diffeomorphisms of the manifold $M$, and the
rescaled metric $ \tilde{g}_{\mu\nu}$ is not simply the metric $g_{\mu\nu}$
written in a different coordinate system: the metrics $
\tilde{g}_{\mu\nu}$ given by Eq.~(\ref{CT}) and the metric
$g_{\mu\nu}$ describe different gravitational fields and different
physics. Special conformal
transformations originating from diffeomorphisms are called {\em conformal
isometries} (Wald 1984). The reader should not be confused by
the fact that some authors use the name ``conformal transformation'' for
special coordinate transformations relating inertial and accelerated
observers
(e.g. Fulton, Rorlich and Witten
1962{\em a,b}; Wood, Papini and Cai 1989;
Mashoon 1993). In this case the metric is left unchanged,
although its coordinate representation varies.
The possibility of different conformal rescalings for different metric
components has also been considered (Mychelkin 1991), although it appears
doubtful that this procedure can be given a covariant formulation and a
physically sound motivation.

Historically, interest in conformal transformations arose after the
formulation of Weyl's (1919) theory
aimed at unifying gravitation and electromagnetism, expecially after its
reformulation by Dirac (1973). Moreover, a conformally
invariant version of special relativity was formulated
(Page 1936{\em a,b}; Page and Adams 1936), but the conformal invariance in
this case was recognized to be meaningless (Pauli 1958). Further
developments of Weyl's theory are more appealing; for example, the
self--consistent, scale--invariant theory of Canuto {\em et al.} (1977),
so far, is not in contraddiction with the observations. It requires that
the astronomical unit of length is related to the atomic unit by a scalar
function which depends on the spacetime point. The theory contains a
time--dependent cosmological ``constant'' $ \Lambda (t)=\Lambda_0
(t_0/t )^2 $, which is sought after by many authors in modern cosmology and
astroparticle physics. 

\section{Conformal transformations as a mathematical tool}

\setcounter{equation}{0}
Conformal rescalings and conformal techniques have been widely used in
general
relativity for a long time, expecially
in the theory of asymptotic flatness and in the initial value formulation
(Wald 1984
and references therein), and also in studies of the
propagation of massless fields,
including Fermat's principle (Perlick 1990; Schneider, Ehlers and
Falco 1992), gravitational lensing in
the (conformally flat) Friedmann--Lemaitre--Robertson--Walker universe
(Perlick 1990; Schneider, Ehlers and Falco 1992),
wave equations (Sonego and Faraoni 1992; Noonan 1995),
studies of the optical geometry near black hole horizons
(Abramowicz, Carter and Lasota 1988; Sonego and Massar 1996; Abramowicz
{\em et al.} 1997{\em a,b}), exact solutions (Van
den Bergh 1986{\em a,b,c,d,e}, 1988) and in other contexts.
Conformal techniques and conformal invariance are important also for
quantum field
theory in curved spaces (Birrell and Davies 1982), for statistical
mechanics and for
string theories (e.g. Dita and Georgescu 1989).
A conformal transformation is often used as a mathematical tool to map the
equations of motion of physical
systems into mathematically equivalent sets of equations
that are more easily solved and
computationally more convenient to study. This situation arises mainly in
three
different areas of gravitational physics: alternative (including nonlinear)
theories of gravity, unified
theories in multidimensional spaces, and studies of scalar fields
nonminimally coupled to gravity. \\ \\
{\bf Brans--Dicke theory:} The conformal rescaling to the
minimally coupled case for the Brans--Dicke field in Brans--Dicke theory
was
found by Dicke (1962). One starts with the Brans--Dicke action in
the so--called ``Jordan frame''
\be         \label{BDaction}
S_{BD}=\frac{1}{16\pi}\int d^4 x \sqrt{-g} \left[
\phi \, R  -\frac{\omega}{\phi} \, \nabla^{\mu}
\phi \nabla_{\mu} \phi \right] + S_{matter}  \; ,
\ee
which corresponds to the field equations
\be  \label{BD1}
R_{\mu\nu}-\frac{1}{2} g_{\mu\nu} R= \frac{8\pi}{\phi} T_{\mu\nu}
+\frac{\omega}{\phi^2}  \left( \nabla_{\mu} \phi \nabla_{\nu} \phi
-\frac{1}{2} g_{\mu\nu} \nabla^{\alpha} \phi
\nabla_{\alpha} \phi \right) +\frac{1}{\phi} \left(
\nabla_{\mu}\nabla_{\nu}\phi  -g_{\mu\nu} \Box \phi \right) \; ,
\ee
\be \label{BD2}
\Box \Phi =\frac{8\pi T}{3+2\omega} \; .
\ee
The conformal transformation (\ref{CT}) with
\be  \label{OmBD}
\Omega =\sqrt{G \phi}
\ee
and the redefinition of the scalar field given in differential form by
\be   \label{8}
d\tilde{\phi}=\sqrt{ \frac{2\omega +3}{16 \pi G}} \, \, \frac{d\phi}{\phi}
\ee
($\omega > -3/2$) transform the action (\ref{BDaction}) into the
``Einstein frame'' action
\be    \label{actionBDEframe}
S=\int d^4 x \left\{ \sqrt{-\tilde{g}} \left[
\frac{\tilde{R}}{16 \pi G} -\frac{1}{2} \tilde{\nabla}^{\mu}
\tilde{\phi} \tilde{\nabla}_{\mu} \tilde{\phi} \right] + \exp \left( -8
\sqrt{\frac{\pi G}{2\omega+3}}  \tilde{\phi} \right) 
{\cal L}_{matter}( \tilde{g})  \right\} \;  ,
\ee
where $\tilde{\nabla}_{\mu}$ is the covariant derivative operator
of the rescaled metric $ \tilde{g}_{\mu\nu}$.
The gravitational part of the action now contains only Einstein gravity,
but a free scalar field acting as a source of gravitation {\em always}
appears. It permeates spacetime in a way that cannot be eliminated, i.e.
one cannot contemplate solutions of the vacuum Einstein equations
$R_{\mu\nu}=0$ in the Einstein frame.
In the Jordan frame, the gravitational field is described by the metric
tensor
$g_{\mu\nu}$ {\em and} by the Brans--Dicke field $\phi$. In the Einstein
frame, the gravitational field is described only by the metric tensor $
\tilde{g}_{\mu\nu}$, but the scalar field $\tilde{\phi}$, which is now a
form of matter, is always present, a reminiscence of its fundamental role
in the ``old'' frame. In addition, the rest of the matter part of the
Lagrangian is multiplied by an exponential factor, 
thus  displaying an anomalous coupling to the scalar 
$\tilde{\phi}$. This anomalous coupling will be discussed in Sec.~6.\\\\
{\bf Nonminimally coupled scalar field:} By means
of a conformal rescaling, the study of a nonminimally coupled scalar
field can also be reduced to that of a minimally coupled scalar.
The transformation relating a massless conformally coupled and
a minimally coupled scalar field was found by Bekenstein
(1974) and
later rediscovered and generalized to massive fields and arbitrary
values of the coupling constant
(Deser 1984; Schmidt 1988; Maeda 1989; Futamase and Maeda 1989;
Xanthopoulos and Dialynas 1992; Klimcik 1993;
Accioly {\em et al.} 1993). In
this case, the starting point is the action for canonical gravity plus a
scalar field in the Jordan frame:
\be   \label{nonmincoupl}
S=\int d^4 x \sqrt{-g}\left[ \left( \frac{1}{16\pi G} -\frac{\xi \phi^2}{2}
\right) R
-\frac{1}{2} \nabla^{\mu} \phi \nabla_{\mu} \phi -V( \phi ) \right] \; ,
\ee
where $V( \phi) $ is the scalar field potential (possibly
including a mass
term and the cosmological constant) and $\xi$ is a dimensionless coupling
constant. Note that the dimensions of the scalar field are $ \left[ \phi
\right] = \left[ G^{-1/2} \right] = \left[ m_{pl}\right] $. The equation
satisfied by the scalar $\phi$ is 
\be  \label{ABAB}
\Box \phi-\xi R \phi -\frac{dV}{d\phi}=0 \; .
\ee
Two cases occur most frequently in the literature: ``minimal coupling''
($\xi=0$) and ``conformal coupling'' ($\xi=1/6$); the latter makes the
wave equation (\ref{ABAB}) conformally invariant in four dimensions if
$V=0$ or $V=\lambda \phi^4$ (the
latter potential being used in the chaotic inflationary scenario).
The conformal transformation (\ref{CT}) with
\be   \label{OM}
\Omega^2=1-8\pi G \xi \phi^2
\ee
and the redefinition of the scalar field, given in differential form by
\be         \label{redefNMC}
d\tilde{\phi}=\frac{\left[ 1- 8\pi G \xi \left( 1-6\xi \right) \phi^2
\right]^{1/2}} {1- 8\pi G \xi \phi^2} \, d\phi \; ,
\ee
reduce (\ref{nonmincoupl}) to the Einstein frame action
\be   \label{mincoupl}
S=\int d^4 x \sqrt{-\tilde{g}} \left[ \frac{\tilde{R}}{16\pi G}
-\frac{1}{2} \tilde{\nabla}^{\mu} \tilde{\phi}
\tilde{\nabla}_{\mu}\tilde{\phi}
-\tilde{V}( \tilde{\phi} ) \right] \; ,
\ee
where the scalar field $\tilde{\phi}$ is now minimally coupled  and
satisfies the equation
\be
\tilde{g}^{\mu\nu} \nabla_{\mu} \nabla_{\nu} \tilde{\phi} 
-\frac{d\tilde{V}}{d\tilde{\phi}}=0 \; .
\ee
The new scalar field potential is given by
\be
\tilde{V} ( \tilde{\phi})=\frac{V( \phi)}{\left( 1-8\pi G \xi \phi^2
\right)^2} \; ,     
\ee
where $\phi=\phi \left( \tilde{\phi} \right)$ is obtained by integrating
and inverting
Eq.~(\ref{redefNMC}). The field equations of
a gravitational theory in the case of a minimally coupled scalar
field as a source of
gravity are computationally much easier to solve than the corresponding
equations for nonminimal coupling, and the transformation (\ref{CT}),
(\ref{OM}), (\ref{redefNMC}) is widely used for this purpose.
The stress--energy tensor of a scalar field can be put in the form
corresponding to a fluid, but the $T_{\mu\nu} $ for a nonminimally
coupled field is considerably more complicated than the minimal coupling
case, for which the form of the $T_{\mu\nu} $ reduces to that of a perfect
fluid (Madsen 1988). It is generally assumed that the scalar field $\phi$
assumes values in a range that makes the right hand side of
Eq.~(\ref{OM}) positive. For $\xi >0$, this range is limited by the
critical values $\phi_{1,2} =\pm \left( 8\pi G \xi \right)^{-1/2}$.

Nonminimal couplings of the electromagnetic field to gravity have also been
considered
(Novello and Salim 1979; Novello and Heintzmann 1984; Turner and Widrow
1988; Novello and Elbaz 1994; Novello, Pereira and Pinto--Neto 1995;
Lafrance and Myers 1995), but
conformal techniques analogous to those developed for scalar fields are
presently unknown. A formal method alternative to conformal transformations is
sometimes useful
for nonminimally coupled scalar fields, which are equivalent to an
effective
flat space field theory with a scalar mass that is $\xi$--dependent
(Hochberg and Kephart 1995).
\\ \\
{\bf Nonlinear theories of gravity:} The mathematical equivalence between a
theory described by the gravitational Lagrangian density
${\cal L}_g=\sqrt{-g} f(R) $ (``higher order theory'') and
Einstein gravity was found in
(Teyssandier and Tourrenc 1983; Schmidt 1987; Starobinsky
1987; Barrow and Cotsakis 1988; Maeda 1989;
Gott, Schmidt and Starobinsky 1990; Schmidt 1990; Cotsakis
and Saich 1994; Wands 1994). The field equations for this theory are of
fourth order:
\be
\left( \frac{df}{dR} \right) R_{\mu\nu}-\frac{1}{2} \, f(R)g_{\mu\nu}
-\nabla_{\mu}
\nabla_{\nu} \left( \frac{df}{dR} \right) + g_{\mu\nu} \Box \left(
\frac{df}{dR} \right) =0 \; ,
\ee
and are reduced to the Einstein equations by the conformal transformation.

Quadratic Lagrangian densities with $R^2$ terms arising from quantum
corrections are the most frequently studied cases of
nonlinear gravitational theories; they
can be reduced to the Einstein Lagrangian density
(Higgs 1959; Teyssandier and Tourrenc 1983; Whitt 1984;
Ferraris 1986; Berkin and Maeda 1991). These
results were generalized to supergravity, Lagrangians with terms
$\Box^k R$ ($k \geq 1$) and polynomial Lagrangians  in $R$ (Cecotti
1987);
the two--dimensional case was studied in (Mignemi and Schmidt 1995).
This class of theories includes Weyl's theory (Weyl 1919; Dirac 1973)
described by the Lagrangian density $
{\cal L}=\sqrt{-g} ( R^2+\beta F_{\mu\nu}F^{\mu\nu} ) $, and theories of
the form ${\cal L} =R^k$ ($k \geq 1$).
For nonlinear theories of gravity, the conformal transformation that maps the
theory into Einstein gravity becomes a Legendre transformation
(Ferraris, Francaviglia
and Magnano 1988;
Jakubiec and Kijowski 1988; Ferraris,
Francaviglia and Magnano 1990; Magnano, Ferraris
and Francaviglia 1990; Magnano and Sokolowski 1994).

There are obvious advantages in performing this transformation because the
higher order field equations of the nonlinear theory are reduced to the
second order Einstein equations with matter.
One starts with a purely gravitational nonlinear theory described by the
action
\be   \label{nonlin}
S=\int d^m x \sqrt{-g}\left[  F( \phi, R) -\frac{\epsilon}{2}
\nabla^{\mu} \phi \nabla_{\mu} \phi \right] \; ,
\ee
in $m$ spacetime dimensions, where $F( \phi, R )$ is an arbitrary (but
sufficiently regular) function of $\phi$ and $R$, and $\epsilon $ is a free
parameter (normally $0$ or $1$). 

The corresponding field equations (Maeda 1989) are
\begin{eqnarray}
\left( \frac{\partial F}{\partial R} \right) \left( R_{\mu\nu}
-\frac{1}{2} \, g_{\mu\nu} R \right) & = & \frac{\epsilon}{2} \left(
\nabla_{\mu} \phi \nabla_{\nu} \phi
-\frac{1}{2} g_{\mu\nu} \nabla^{\alpha} \phi \nabla_{\alpha} \phi \right)
+ \frac{1}{2} g_{\mu\nu} \left( F-\frac{\partial F}{\partial R} R \right)
\nonumber \\
& & + \nabla_{\mu}
\nabla_{\nu} \left( \frac{\partial F}{\partial R} \right) - g_{\mu\nu}
\Box \left( \frac{\partial F}{\partial R} \right) \; ,
\end{eqnarray}
\be
\epsilon \Box \phi = - \, \frac{\partial F}{\partial \phi} \; .
\ee 
The conformal rescaling (\ref{CT}), where
\be            \label{17}
\Omega^2 =\left[ 16\pi G \left| \frac{\partial F}{\partial R} \right| +
\mbox{constant} \right]^{2/ (m-2)} \; ,
\ee
and the redefinition of the scalar field
\be            \label{18}
\tilde{\phi} = \frac{1}{\sqrt{8\pi G}} \sqrt{ \frac{m-1}{m-2}} \ln \left[
\sqrt{32\pi} G\left| \frac{\partial F}{\partial R} \right| \right]
\ee
(Maeda 1989) reduce the action (\ref{nonlin}) to
\be   \label{lin}
S=\alpha \int d^m x \sqrt{-\tilde{g}}\left[ \frac{\tilde{R}}{16\pi G}
 -\frac{1}{2} \tilde{\nabla}^{\mu} \tilde{\phi} \tilde{\nabla}_{\mu}
\tilde{\phi}
-\frac{\epsilon \alpha}{2} \exp \left[ - \sqrt{8 \pi G \, \frac{m-2}{m-1}}
\, \tilde{\phi} \right] -U( \phi, \tilde{\phi} ) \right]
\ee
where the two scalar fields $\phi $ and $\tilde{\phi}$ appear and
\be
\alpha = \frac{ \partial F/\partial R}{\left| \partial F/\partial R
\right|} \; ,  
\ee
\be
U( \phi, \tilde{\phi})= \alpha \exp \left( -\, \frac{m \sqrt{8\pi G} \tilde{\phi}}
{\sqrt{(m-1)(m-2)}} \right)
\left[ \frac{\alpha}{16\pi G} R( \phi, \tilde{\phi}) \exp\left(
\sqrt{\frac{m-2}{m-1} 8\pi G}\, \tilde{\phi} \right) -F( \phi, \tilde{\phi})
\right] \; ,
\ee
and $ F( \phi, \tilde{\phi})= F( \phi, R( \phi,\tilde{\phi})) $.
The resulting system is of
nonlinear $\sigma$--model type, canonical gravity with two scalar fields
$\phi$, $\tilde{\phi} $.

In the particular case in which $F( \phi, R)$ is a linear function of the
Ricci curvature,
\be
F( \phi, R)=f( \phi) R -V( \phi) \; ,
\ee
the redefinition of the scalar field
\be  \label{22}
\tilde{\phi}=\frac{1}{\sqrt{8 \pi G}} \int d\phi \left\{ \frac{\epsilon
(m-2) f( \phi)
+2(m-1) \left[ df( \phi) /d\phi \right]^2}{2(m-2)f^2( \phi)}
\right\}^{1/2}
\ee
(where the argument of the square root is assumed to be positive) leads to
the
Einstein action with a single scalar field $\phi$:
\be      \label{actionEframe}
S=\frac{|f|}{f} \int d^m x \sqrt{-\tilde{g}}\left[ \frac{\tilde{R}}{16\pi
G}-\frac{1}{2} \tilde{\nabla}^{\mu} \tilde{\phi}
\tilde{\nabla}_{\mu} \tilde{\phi}  -U( \tilde{\phi} ) \right] \; .
\ee
This action is equivalent to the Einstein equations
\be
\tilde{R}_{\mu\nu}-\frac{1}{2} \tilde{g}_{\mu\nu} \tilde{R}=8 \pi G
\, \tilde{T}_{\mu\nu}
\left[ \tilde{\phi} \right] \; ,
\ee
\be
\tilde{T}_{\mu\nu} \left[ \tilde{\phi} \right] = \nabla_{\mu} \tilde{\phi}
\nabla_{\nu} \tilde{\phi} -\frac{1}{2} \,
\tilde{g}_{\mu\nu}\tilde{g}^{\alpha\beta} 
\nabla_{\alpha} \tilde{\phi}\nabla_{\beta} \tilde{\phi}+ U
\, \tilde{g}_{\mu\nu} \; ,
\ee
where
\be
U( \tilde{\phi})= \frac{|f|}{f} \left[ 16\pi G \left| f( \phi) \right|
 \right]^{\frac{-m}{m-2}} \, V( \phi )
\ee
and $\phi =\phi \left( \tilde{\phi} \right) $.
The transformations (\ref{OmBD}), (\ref{OM}) and (\ref{redefNMC}) are
recovered
as particular cases of (\ref{22}), (\ref{17}).
In addition, all the theories described by a four--dimensional action of
the form
\be
S=\int d^4 x \sqrt{-g} \left[ f( \phi ) R +A( \phi) {\nabla}^{\mu} \phi
{\nabla}_{\mu} \phi +V( \phi ) \right]
\ee
and satisfying the relation
\be
2Af-3\left( \frac{df}{d\phi} \right)^2 =0\; , \;\;\;\;\;\;\; V(
\phi)=\lambda
f^2 ( \phi)
\ee
($\lambda=$constant) are conformally related (Shapiro and Takata 1995);
particular
cases include general relativity and the case of a conformally coupled
scalar
field.

The conformal transformation establishes a {\em mathematical} equivalence
between the theories formulated in the two conformal frames;
the space of solutions of the theory in one frame is isomorphic to the
space of
solutions in the conformally related frame (which is mathematically more
convenient to study). The conformal transformation can also be used as a
solution--generating technique, if solutions are known in one conformal
frame but not in another
(Harrison 1972;
Belinskii and Kalatnikov 1973;
Bekenstein 1974;
Van den Bergh 1980, 1982, 1983{\em a,b,c,d};
Froyland 1982;
Accioly, Vaidya and Som 1983;
Lorentz--Petzold 1984;
Barrow and Maeda 1990;
Klimcik and Kolnik 1993;
Abreu, Crawford and Mimoso 1994).
It is to be stressed that the mathematical equivalence
between the two systems {\em a priori} implies nothing about
their physical equivalence
(Brans 1988; Cotsakis 1993; Magnano and Sokolowski 1994). Moreover,
only the gravitational
(vacuum) part of the action is conformally equivalent to Einstein gravity:
if ordinary matter
(i.e. matter different from the scalar field used in the conformal
transformation) is added to the theory, the coupling of this matter to
gravity and the conservation equations that it satisfies
are different in the two conformally related frames.
The advantage of the conformal transformation in a
non--purely vacuum theory is questionable: it has
been argued that, because the Einstein frame scalar field is
coupled to matter, a simplification of the equations
of motion in this case does not occur (Barrow and Maeda 1990).

Not only it is possible to map the classes of theories considered above
into
canonical Einstein gravity, but it is also possible to find conformal
transformations between each two of these theories (see
Magnano and Sokolowski 1994 for a table of possible transformations).
Indeed, one expects to be able to do that by
taking appropriate compositions of different maps from gravitational
theories to general relativity, and their inverse maps.

We conclude this section with a remark on the terminology: it has become
common to use the word ``frame'' to
denote a set of dynamical variables of the theory considered;
the term ``gauge'' instead of ``frame'' has been
(rather improperly) used (Gibbons and Maeda 1988; Brans 1988). In
some papers
(Cho 1987, 1990, 1992, 1994, 1997; Cho and Yoon 1993)
the metric $\tilde{g}_{\mu\nu} $ in the Einstein frame is called
``Pauli metric'', as opposed to the ``Jordan'' or ``atomic unit'' metric
$g_{\mu\nu} $ of the Jordan frame.

\section{Is the Einstein frame physical~?}

\setcounter{equation}{0}
Many high energy theories and many classical gravity theories are
formulated by
using a conformal transformation mapping the Jordan 
frame to the Einstein frame. Typically, the conformal factor of
the transformation is a function of a dilaton, or
Brans--Dicke--like field already present in the theory.
The classical theories of gravity for which a conformal transformation maps
the system into a new conformal frame, in which the gravitational sector
of the
theory reduces to the canonical Einstein form,
include Brans--Dicke theory and its scalar--tensor
generalizations, non--linear gravity theories, classical Kaluza--Klein
theories
and in general, all theories which have an extended gravitational sector or
which involve a dimensional reduction and compactification of extra
spacetime
dimensions. Quantum theories incorporating the conformal transformation
include
superstring and supergravity theories and $\sigma$--models. The
transformation
to the Einstein frame seems to be universally accepted for
supergravity and superstring theories (although field redefinitions may be
an issue for debate (Tseytlin 1993)). It is unknown
whether physics is
conformally invariant at a sufficiently high energy scale, but there are
indications in this sense from string theories (Green, Schwarz and Witten
1987) and
from $SU(N)$ induced gravity models in which, in the high energy limit,
the scalar fields of the theory approach conformal coupling
(Buchbinder, Odintsov and Shapiro 1992; Geyer and Odintsov 1996).
We have no experiments capable of probing the energy scale of string
theories, and
conformal invariance at this energy scale cannot be directly tested.
While the low--energy Einstein gravity
contains a dynamical degree of freedom connected with the ``length'' of the
metric tensor (the determinant $g$), this is absent in conformally
invariant gravity (e.g. induced gravity described by the action
(\ref{inducedgravity})).
The conformal invariance of a theory implies that the latter contains no
intrinsic mass; a nonzero mass would introduce a preferred length scale in
the theory, 
thus breaking the scale--invariance.
The physical inequivalence of conformal
frames at low energies reflects the fact that the non--negligible
masses of the fields present in the theory break the conformal symmetry
which
is present at higher energies. In classical gravity theories,
there is disagreement and confusion
on the long--standing (Pauli 1955; Fierz 1956)
problem of which conformal frame is the physical one. Is the Jordan frame
physical and the Einstein frame unphysical~? Is the conformal
transformation necessary, and the Einstein frame physical~?
Does any other choice of the conformal factor in Eq.~(\ref{CT}) map the
theory
into a physically significant frame, and how many of these theories are
possible~? Here the term
``physical'' theory denotes one that is theoretically consistent and
predicts the values of some observables that can, at least in principle,
be measured in experiments performed in four macroscopic spacetime
dimensions (definitions that differ from ours are
sometimes adopted in the literature, see  e.g. (Garay
and Garcia--Bellido 1993; Overduin and Wesson 1997)). The ambiguity in the
choice of the physical conformal frame raises also problems
of an almost philosophical character (Weinstein 1996).

Before attempting to answer any of these questions, it is important
to recognize that, in general, the
reformulation of the theory in a new conformal frame leads to a different,
physically inequivalent theory.
If one restricts oneself to consider the metric tensor and physics that
does
not involve only conformally invariant fields (e.g. a stress--energy tensor
$T_{\mu\nu} $ with nonvanishing trace), or experiments involving massive
particles and timelike observers, it is obvious that metrics conformally
related by a nontrivial transformation of the kind (\ref{CT}) on a manifold
describe different gravitational
fields and different physical situations. For example, one could consider a
Friedmann--Lemaitre-Robertson--Walker metric with flat spatial sections,
given
by the line element
\be   \label{EdS}
ds^2=a^2 ( \eta ) \left( -d\eta^2+dx^2+dy^2+dz^2 \right) \; ,
\ee
where $\eta$ is the conformal time and $(x,y,z)$ are spatial comoving
coordinates. The metric (\ref{EdS}) is conformally flat, but certainly it is
not
physically equivalent to the Minkowski metric $\eta_{\mu\nu}$, since it
exhibits a nontrivial dynamics and significant (observed) cosmological
effects. 

The authors working in classical gravitational physics can be grouped
into five categories according to their attitude towards the issue of the
conformal frame (we partially follow a previous classification by
Magnano and Sokolowski (1994)):
\begin{itemize}

\item authors that neglect the issue
(Deruelle and Spindel 1990;
Garcia--Bellido and Quir\'{o}s 1990; Hwang 1990;
Gottl\"{o}ber, M\"{u}ller and Starobinsky 1991;
Suzuki and  Yoshimura 1991;
Rothman and Anninos 1991;
Guendelman 1992;
Guth and Jain 1992;
Liddle and Wands 1992;
Capozziello, Occhionero and Amendola 1993;
Capozziello, de Ritis and Rubano 1993;
McDonald 1993{\em a,b};
Barrow, Mimoso and de Garcia Maia 1993;
Garcia--Bellido and Wands 1995;
Laycock and Liddle 1994;
Alvarez and Bel\'{e}n Gavela 1983;
Sadhev 1984;
Deruelle and Madore 1987;
Van den Bergh and Tavakol 1993;
Fabris and Sakellariadou 1997;
Kubyshin and Martin 1995;
Fabris and Martin 1993;
Chatterjee and Banerjee 1993;
Biesiada 1994;
Liddle and Lyth 1993;
Hwang 1996);

\item authors that explicitely support the view that a theory
formulated in one conformal frame is physically equivalent to the
reformulation of the same theory in a different conformal frame
(Buchm\"{u}ller and Dragon 1989; Holman, Kolb and
Wang 1990; Campbell, Linde and Olive 1991;
Casas, Garcia--Bellido
and Quir\'{o}s 1991; Garay
and Garcia--Bellido 1993; Levin 1995{\em
a,b}; Shapiro and Takata 1995; Kaloper and Olive 1998);

\item authors that are aware of the physical non--equivalence of
conformally
related frames but
do not  present conclusive arguments in favour of one or the other of the
two versions of the theory (and/or perform computations both in the Jordan
and the Einstein frame)
(Brans 1988; Jakubiec
and Kijowski 1988; Kasper and Schmidt 1989; Deruelle and Spindel
1990; Hwang 1990; Kolb, Salopek and Turner 1990; Gottl\"{o}ber,
M\"{u}ller and Starobinsky 1991;
Suzuki and
Yoshimura 1991; Rothman and Anninos
1991; Guendelman 1992; Guth and Jain 1992; Liddle and Wands 1992;
Piccinelli,
Lucchin and
Matarrese 1992; Damour and Nordvedt 1993{\em a,b}; Cotsakis
and Saich 1994; Hu, Turner and Weinberg 1994;
Turner 1993; Mimoso and Wands
1995{\em b}; Faraoni 1996{\em a}; Weinstein 1996; Turner and Weinberg
1997; Majumdar
1997; Capozziello, de Ritis and Marino 1997;
Dick 1998);

\item authors that identify the Jordan frame as physical (possibly allowing
the use of the conformal transformation as a purely mathematical tool)
(Gross and Perry 1983; Barrow and
Maeda 1992; Berkin, Maeda and Yokoyama 1990; 
Damour, Gibbons and Gundlach 1990; 
Kalara, Kaloper and Olive 1990;
Berkin and Maeda 1991;
Damour and Gundlach 1991; Holman, Kolb and Wang 1991;  
Mollerach and Matarrese 1992; Tao
and Xue 1992; Wu 1992; del Campo 1992; Tkacev 1992;
Mignemi and Whiltshire 1992; Barrow 1993; Bruckman and Velazquez 1993;
Cotsakis and Flessas 1993; Will and Steinhardt 1995; Scheel, Shapiro and
Teukolsky 1995; Barros and Romero 1998);

\item authors that identify the Einstein frame as the physical one
(Van den Bergh 1981, 1983{\em e}; Kunstatter, Lee and Leivo 1986;
Gibbons and Maeda 1988; Sokolowski
1989{\em a,b}; Pimentel and Stein--Schabes 1989; Kubyshin, Rubakov
and Tkachev 1989;
Salopek, Bond and
Bardeen 1989; Kolb, Salopek and Turner 1990; Cho
1990; Deruelle, Garriga and Verdaguer 1991; Cho 1992;
Amendola {\em et al.} 1992; 
Amendola, Bellisai and
Occhionero 1993; Cotsakis 1993; Cho and Yoon 1993; Alonso {\em et al.} 1994;
Magnano and Sokolowski
1994; Cho 1994;  Occhionero and Amendola 1994;
Lu and Cheng 1996; Fujii 1998;
Cho 1997; Cho and Keum 1998).

\end{itemize}
Sometimes, works by
the same author(s) belong to two different groups; this illustrates the
confusion on the issue that is present in the literature.

The two conformal frames, however, are substantially different.
Furthermore, if a preferred conformal
frame does not exist, it is possible to generate an infinite number of
alternative theories and of cosmological inflationary scenarios by
arbitrarily choosing the conformal
factor (\ref{specialCT}) of the transformation (\ref{CT}). Only
when a physical frame is
uniquely determined the theory and
its observable predictions are meaningful.

Earlier attempts to solve the problem in Brans--Dicke theory
advocated the equivalence principle: to
this end it is essential to consider not only the gravitational, but also
the matter part of the Lagrangian. The use of the equivalence principle
requires a careful analysis (Brans 1988; Magnano and Sokolowski 1994); by
including the Lagrangian for ordinary matter in the
Jordan frame action, one finds that, after the conformal transformation has
been
performed, the scalar field in the Einstein frame couples minimally to
gravity,
but nonminimally to matter (``non--universal coupling''). Historically,
the Jordan frame was selected as physical because the dilaton couples
minimally to ordinary matter in this frame (Brans and Dicke 1961).
Attempts were also made to derive conclusive results
from the conservation laws for the
stress--energy tensor of matter, favouring the Jordan frame (Brans 1988)
or
the Einstein frame (Cotsakis 1993; Cotsakis 1995 -- see
(Teyssandier 1995; Schmidt 1995; Magnano and Sokolowski 1994) for
the correction of a flaw in the proof of
(Cotsakis 1993; Cotsakis 1995)). Indeed,
the conservation laws do not allow one to
draw definite conclusions (Magnano and Sokolowski 1994).

However, the point of view that selects the Jordan frame as physical
is untenable because it leads to a negative definite, or indefinite
kinetic energy for the scalar field; on the contrary,
the energy density is positive definite in the Einstein frame. This
result was initially proved for Brans--Dicke and for Kaluza--Klein
theories, and later generalized to gravitational theories with Lagrangian
density ${\cal L}=f(R) \sqrt{-g}
+{\cal L}_{matter}$ (Magnano and Sokolowski 1994). This
implies that the theory does not have a stable ground state, and
that the system decays into a lower energy state {\em ad infinitum}
(Gross and Perry 1983; Appelquist and Chodos 1983; Maeda 1986{\em
b}; Maeda 1987).
While a stable ground state may not be required for certain particular
solutions of the theory (e.g. cosmological solutions (Padmanabhan
1988)), or
for Liouville's theory (D'Hoker and Jackiw 1982), it is certainly
necessary for a viable theory of classical gravity. The ground state of
the system must
be stable against
small fluctuations and not fine--tuned, i.e. nearby solutions of the theory
must have similar properties
(Strater and Wightman 1964; Epstein, Glaser and Jaffe 1965; Abbott
and Deser 1982). The fact that the
energy is not positive definite is usually associated with the formulation
of
the theory in unphysical variables. On the contrary, the energy conditions
(Wald 1984) are
believed to be satisfied by all classical matter and fields (not so in
quantum theories~--~see Sec.~8). This decisive
argument was first used to select the Einstein frame in Kaluza--Klein and
Brans--Dicke theories
(Bombelli {\em et
al.} 1987; Sokolowski and Carr 1986;
Sokolowski 1989{\em a,b}; Sokolowski and Golda 1987; Cho 1990; Cho
1994), and later generalized to scalar--tensor theories
(Cho 1997) and nonlinear gravity theories (Magnano and Sokolowski 1994).
Also, the uniqueness of a physical conformal frame was proved
(Sokolowski 1989{\em a,b}; Magnano and Sokolowski 1994).

For completeness, we mention other arguments supporting the Einstein
frame as physical that have appeared in the literature: however, they are
either highly questionable (sometimes to the point of not being valid), 
or not as compelling as the one based on the positivity of
energy. The Hilbert and the Palatini actions for scalar--tensor theories are
equivalent in the Einstein but not in the Jordan frame (Van den Bergh 1981,
1983{\em e}). Some authors choose the Einstein frame on the basis
of the resemblance of its
action with that of general relativity (Gibbons and Maeda
1988; Pimentel and Stein--Schabes
1989; Alonso {\em et al.} 1994; Amendola, Bellisai and Occhionero 1993);
others (Salopek, Bond and Bardeen 1989; Kolb, Salopek and Turner 1990)
find difficulties in quantizing the scalar field fluctuations in the linear
approximation in the Jordan frame, but not in the Einstein frame;
quantization and the conformal transformation do not commute
(Fujii and Nishioka 1990; Nishioka and Fujii 1992; Fakir and Habib 1993). 
Other authors claim
that the Einstein frame is forced upon us by the compactification of the
extra dimensions in higher dimensional theories
(Kubyshin, Rubakov and Tkachev 1989; Deruelle, Garriga and Verdaguer 1991).

A possible alternative to the Einstein frame formulation of the complete
theory (gravity plus matter) has been supported (Magnano and Sokolowski
1994), and consists in starting with the  introduction of  matter
non--minimally  coupled to the Brans--Dicke scalar in the Jordan frame,
with
the coupling tuned in such a way that the Einstein frame action exhibits
matter minimally coupled to the Einstein frame scalar field, after the
conformal
transformation has been performed. This procedure arises
from the observation (Magnano and Sokolowski 1994) that the traditional
way of prescribing
matter minimally coupled in the Jordan frame relies on the implicit
assumptions that \\
{\em i)} the equivalence principle holds;\\
{\em ii)} the Jordan frame is the physical one. \\
While these assumptions are not justified {\em a priori}, as noted
by Magnano and Sokolowski (1994), the possibility of adding matter
in the Jordan frame with a
coupling that exactly balances the exponential factor appearing in the
Einstein frame
appears to be completely {\em ad hoc} and is not physically motivated;
by proceeeding along these lines, one could arbitrarily change the theory
without theoretical justification.

As a summary, the Einstein frame is the physical one (and the Jordan frame
and
all the other conformal frames are unphysical) for the following classes of
theories:
\begin{itemize}

\item scalar--tensor theories of gravity described by the Lagrangian
density
\be
{\cal L}=\sqrt{-g} \left[ f( \phi) R-\frac{\omega ( \phi )}{\phi}
\nabla^{\mu}
\phi \nabla_{\mu} \phi +\Lambda ( \phi) \right]
+{\cal L}_{matter} \; ,
\ee
which includes Brans--Dicke theory as a special case (see Sec. 4 for the
corresponding field equations);

\item classical Kaluza--Klein theories;

\item nonlinear theories of gravity whose gravitational part is described
by the Lagrangian density $ {\cal L}=\sqrt{-g} f(R) $ (see Sec. 4 for the
corresponding field equations).

\end{itemize}
Since the Jordan frame  formulation of alternative theories of gravity is
unphysical, one reaches the conclusion that the Einstein frame formulation
is
the only possible one for a classical  theory. In other words, this amounts
to
say that Einstein gravity is essentially the only viable classical
theory of gravity (Bicknell 1974; Magnano and Sokolowski 1994; Magnano
1995; Sokolowski 1997). We remark that this statement is strictly correct
only if the purely gravitational part of the action (without matter) is
considered: in fact, when matter is included into the action, in general
it exhibits an anomalous coupling to the scalar field which does not
occur in general relativity.

Finally, we comment on the case of a nonminimally coupled scalar field
described by the action (\ref{nonmincoupl}). From the above discussion,
one may be induced to believe that the
Einstein frame description is necessary also in this case: this conclusion
would be incorrect because the kinetic term in the action
(\ref{nonmincoupl})
is
canonical and positive definite, and the problem discussed above for
other theories
of gravity of the form ${\cal L}=\sqrt{-g} f(R)$ does not arise.
It is, however, still true that the Einstein and the Jordan frame are
physically inequivalent: the conformal transformation (\ref{CT}),
(\ref{OM}), (\ref{redefNMC}) implies only a
mathematical, not a physical equivalence, despite strong statements on this
regard that point to the contrary (Accioly {\em et al.} 1993).

\section{Conformal transformations in gravitational theories}

\setcounter{equation}{0}
In this section, we review in greater detail the arguments that led to the
conclusions of the previous section, devoting more attention to specific
classical theories of gravity.\\ \\
{\bf Brans--Dicke theory:} The Jordan--Fierz--Brans--Dicke theory
(Jordan 1949; Jordan 1955; Fierz 1956; Brans and Dicke 1961) described by
the action (\ref{BDaction}) (where $\phi $ has the dimensions of the
inverse gravitational constant,
$ \left[ \phi \right] =\left[ G^{-1} \right] $)
has been the subject of renewed interest, expecially in cosmology in the
extended inflationary scenario
(La and Steinhardt 1989; Kolb, Salopek and Turner 1990; Laycock and Liddle
1994). The recent surge of interest appears to be motivated by a
restricted
conformal invariance of the gravitational part of the
Lagrangian that mimics the conformal invariance of string theories before
conformal symmetry is broken (Cho 1992; Cho 1994; Turner 1993; Kolitch and
Eardley 1995; Brans 1997; Cho and Keum 1998).
The conformal transformation (\ref{CT}) with 
$\Omega=\left( G\phi \right)^{\alpha}$,
together with the redefinition of the scalar field
$\tilde{\phi}=G^{-2\alpha} \phi^{1-2\alpha}$ ($\alpha \neq 1/2$)
maps the Brans--Dicke action (\ref{BDaction}) into an action of
the same form, but with parameter
\be    \label{omegatilde}
\tilde{\omega}=\frac{\omega -6\alpha \left( \alpha -1 \right)}{\left(
1-2\alpha \right)^2} \; .
\ee
If $\omega=-3/2 $, the action (\ref{BDaction}) is invariant under the
conformal
transformation; this case corresponds to the singularity $\alpha
\rightarrow -1/2 $ in the expression (\ref{omegatilde}), but the field
equations (\ref{BD1}), (\ref{BD2}) are not defined in this case.
This conformal invariance is broken when a term describing
matter with $T \equiv {T^{\mu}}_{\mu} \neq  0$
is added to the purely gravitational part of the Brans--Dicke
Lagrangian. This property of conformal invariance of the
gravitational sector of the theory is enjoyed also by a subclass of more
general tensor--multiscalar theories of gravity (Damour and
Esposito--Far\`{e}se 1992) and has not yet been investigated in depth in the
general case. The study of the conformal invariance property of
Brans--Dicke theory helps to solve the problems arising in the 
$\omega \rightarrow \infty $ limit of Brans--Dicke theory (Faraoni 1998).
This limit is supposed to give back general relativity, but it fails to do
so when $T=0$. The differences between the Jordan
and the Einstein frame formulations of Brans--Dicke theory
have been pointed out clearly in (Guth and Jain 1992). It has been noted
in studies of gravitational collapse to a black hole in Brans--Dicke
theory
that the noncanonical form of the Brans--Dicke scalar energy--momentum
tensor in the Jordan frame violates the null energy
condition ($ R_{\alpha\beta} l^{\alpha} l^{\beta}\geq 0 $ for all null
vectors $ l^{\alpha}$). This fact is responsible for a {\em decrease} in
time of the horizon area during the dynamical phase of the collapse, 
contrarily to the case of general relativity (Scheel, Shapiro and
Teukolsky 1995).
The violation of the weak energy condition in the Jordan frame has also
been pointed out (Weinstein 1996; Faraoni and Gunzig 1998{\em a}).

Brans--Dicke theory must necessarily be
reformulated in the Einstein frame; the strongest argument supporting
this conclusion is obtained by observing that the kinetic energy term for
the Brans--Dicke field in the Jordan frame Brans--Dicke Lagrangian
does not have the canonical form for a scalar field, and it is negative
definite
(Gross and Perry 1983; Appelquist and Chodos
1983; Maeda 1986{\em b}; Maeda 1987; Sokolowski and Carr 1986;
Maeda and Pang 1986;
Sokolowski 1989{\em a,b}; Cho 1992, 1993; Magnano and Sokolowski
1994).
The fact that this energy argument was originally developed for
Brans--Dicke and for Kaluza--Klein theories is
not surprising, owing to the fact that
Brans--Dicke theory can be derived from a Kaluza--Klein
theory with $n$ extra dimensions and Brans--Dicke parameter
$\omega=-(n-1) / n $
(Jordan 1959; Brans and Dicke 1961; Bergmann 1968; Wagoner 1970;
Harrison 1972;
Belinskii and Kalatnikov 1973; Freund 1982; Gross and Perry 1983; Cho
1992); this derivation is
seen as a motivation for Brans--Dicke theory, and provides a useful way
of generating exact solutions in one theory from known solutions in the
other (Billyard and Coley 1997). Despite this derivation from
Kaluza--Klein theory, the Jordan frame Brans--Dicke theory is sometimes
considered in $ D>4 $ spacetime dimensions (e.g. Majumdar 1997).

An independent argument supporting the choice of the Einstein
frame as the
physical one is obtained by considering (Cho 1992; Damour and
Nordvedt 1993{\em a,b})
the linearized version of the theory. In the Jordan frame the metric is
$\gamma_{\mu\nu}=\eta_{\mu\nu} +h_{\mu\nu} $ (where $\eta_{\mu\nu} $ is
the Minkowski metric), while in the Einstein frame the conformally
transformed metric is 
\be
\tilde{\gamma}_{\mu\nu}=\gamma_{\mu\nu}
\exp \left( \sqrt{\frac{16\pi G}{2\omega +3}} \, \tilde{\phi} \right) 
\simeq \eta_{\mu\nu} +\rho_{\mu\nu} \; , 
\ee
where
\be
\rho_{\mu\nu}=h_{\mu\nu} + \left( \sqrt{ \frac{16\pi G}{2\omega +3}} \,
\tilde{\phi}  \right)  \eta_{\mu\nu} \; . 
\ee
The canonical action for a spin~2 field is not obtained from
the metric $h_{\mu\nu}$, but it is instead given by $\rho_{\mu\nu}$, and
the
spin~2 gravitational field is described by the Einstein frame corrections
$\rho_{\mu\nu} $ to the flat metric. The Jordan frame corrections
$h_{\mu\nu}$
to $\eta_{\mu\nu}$ describe a mixture of spin~0 and spin~2 fields (the
fact that spin~0 and spin~2 modes are mixed together can also be seen from
the full equations of motion of the theory).

A third argument has been proposed against the choice of the Jordan frame
as the physical one: when quantum corrections are taken into account,
one cannot
maintain the minimal coupling of ordinary (i.e. other than the dilaton)
matter to the Jordan metric (Cho 1997). This nullifies the
traditional
statement that the Jordan frame is to be preferred because the scalar
couples
minimally to all forms of matter in this frame.

These results are of the outmost importance for the experiments aimed at
testing Einstein's theory: the Jordan frame versions of alternative
classical theories of gravity are simply nonviable. However,
despite the necessity of
formulating Brans--Dicke theory in the Einstein frame, the
classical
tests of gravity for this theory are studied only for the Jordan frame
formulation. In general, the authors working on the
experimental tests of general relativity and alternative gravity theories
do
not seem to be aware of this paradoxical situation (e.g.
Reasenberg {\em et al.} 1979; Will 1993).

The conformal rescaling has been used as a mathematical technique to
generate exact 
solutions of Brans--Dicke theory from known solutions of the Einstein
equations (Harrison 1972; Belinskii and Kalatnikov 1973; Lorentz--Petzold
1984) and approximate solutions in the linearized theory (Barros and Romero 
1998). \\\\
{\bf (Generalized) scalar--tensor theories:} This class of theories
(Bergmann 1968; Wagoner 1970; Nordvedt 1970; Will 1993) is described by
the Lagrangian density
\be   \label{nonlin2}
S= \int d^4 x \sqrt{ -g}\left[ f( \phi ) R -\frac{\omega}{2}
\nabla^{\alpha} \phi \nabla_{\alpha} \phi -V( \phi ) \right] +S_{matter} \; ,
\ee
where $\omega=\omega( \phi)$ and
$V =V ( \phi)$ (or by the more
general action (\ref{nonlin})). The corresponding field equations are
\begin{eqnarray}
f( \phi) \left( R_{\mu\nu} -\frac{1}{2} g_{\mu\nu} R \right) & = &
\frac{1}{2}\,  
T_{\mu\nu} + \frac{\omega}{2} \left( \nabla_{\mu} \phi \nabla_{\nu} \phi -\frac{1}{2}
\, g_{\mu\nu} \nabla^{\alpha} \phi \nabla_{\alpha} \phi \right) 
\nonumber  \\ 
& & +\frac{1}{2}\, g_{\mu\nu} \left[ Rf( \phi) -2V \right] +
 \nabla_{\mu} \nabla_{\nu} f -g_{\mu\nu} \Box f \; ,
\end{eqnarray}
\be
\Box \phi +\frac{1}{\omega}\left( \frac{1}{2} \, \frac{ d\omega}{d\phi}
\nabla^{\alpha}\phi \nabla_{\alpha}\phi + \frac{df}{d\phi} - \frac{d
V}{d\phi}\right)   =0  \; ,
\ee
where $T_{\mu\nu} = 2 (-g )^{-1/2} \delta {\cal L}_{matter}/\delta
g^{\mu\nu}$. The action
(\ref{nonlin2}) contains Brans--Dicke
theory (\ref{BDaction}) and the nonminimally coupled scalar field theory
(\ref{nonmincoupl}) as particular cases. Theories with more
than one scalar field have also been investigated
(Damour and Esposito--Far\`{e}se 1992; Berezin {\em et al.} 1989; Rainer
and Zuhk 1996).
A revival of interest in scalar--tensor theories
was generated by the fact that
in supergravity and superstring theories, scalar fields are associated to
the
metric tensor field, and that a coupling between a
scalar field and gravity seems unavoidable in string theories
(Green, Schwarz and Witten 1987). Indeed, scalar fields have been present in
relativistic gravitational theories even before general relativity was
formulated (see Brans 1997 for an historical perspective).

The necessity of
the conformal transformation to the Einstein frame has been advocated in
(Cho 1992, 1997) by investigating which linearized
metric describes the physical spin~2
gravitons. A similar argument was presented in
(Damour and Nordvedt 1993{\em a,b}),
although these authors did not see it as a
compelling reason to select the
Einstein frame as the physical one. It has also been pointed out
(Teyssandier and Tourrenc 1983; Damour and
Esposito--Far\`{e}se 1992; Damour and Nordvedt 1993{\em a,b})
that the mixing of
$g_{\mu\nu}$
and $ \phi$ in the Jordan frame equations of motion
makes the Jordan frame
variables an inconvenient set for formulating the Cauchy problem.
Moreover, the generalization to the case of tensor--multi scalar theories
of gravitation, where several scalar fields instead of a single one
appear, is straightforward
in the Einstein frame but not so in the Jordan frame (Damour and
Esposito--Far\`{e}se 1992). In the Einstein frame, ordinary matter does not
obey the conservation law
$\tilde{\nabla}^{\nu} \tilde{T}_{\mu\nu}=0$ (with the exception of a
radiative fluid with $\tilde{T}=0$, which is conformally invariant)
because of the coupling to the dilaton $\phi$. Instead, the equation
\be  \label{qq}
\tilde{\nabla}_{\nu} \tilde{T}^{\mu\nu}=-\frac{1}{\Omega}
\frac{\partial \Omega}{\partial \phi} \,
\tilde{T} \, \tilde{\nabla}^{\mu}\phi
\ee
is satisfied. The total energy--momentum tensor of matter plus the
scalar field is conserved (see Magnano and Sokolowski 1994 for a detailed
discussion of conservation laws in both conformal frames).

The phenomenon of the propagation of light through scalar--tensor
gravitational waves and the resulting time--dependent amplification
of the light source provide an example of the physical difference between
the Jordan and the Einstein frame.. In the Jordan frame the amplification
effect is of first order in the gravitational wave amplitude (Faraoni
1996{\em a}), while it is only of second order in the Einstein frame
(Faraoni and Gunzig 1998{\em a}).

It is interesting to note that, while the observational constraints on the
Brans--Dicke parameter $\omega$ is $\omega > 500 $ (Reasenberg {\em et
al.} 1979),
Brans--Dicke theory in the Einstein frame is subject to the much more
stringent
constraint $ \omega > 10^8 $ (Cho 1997). However, since the Einstein
frame is the physical one, it is not very meaningful to present constraints
on the Jordan frame parameter $\omega $.
Other formal and physical differences occur in the Jordan and the Einstein
frame: the singular points $\omega \rightarrow \infty$ in the
$\omega$--parameter space of
the Jordan frame correspond to a minimum of the coupling factor $\ln
\Omega(
\phi) $ in the Einstein frame (Damour and Nordvedt 1993{\em a,b}).
Singularities of the
scalar--tensor theory may be smoothed out in the Jordan frame, but they
reappear in the Einstein frame and plague the theory again due to
the fact that the kinetic terms are canonical and the energy conditions
(which are crucial
in the singularity theorems) are satisfied in the Einstein frame
(Kaloper and Olive 1998).

In (Bose and Lohiya 1997), the quasi--local mass defined in general
relativity by the recent Hawking--Horowitz prescription (Hawking and
Horowitz 1996) was generalized to $n$--dimensional scalar--tensor
theories. It was shown that this quasi--local mass is invariant under the
conformal transformation that reduces the gravitational part 
of the scalar--tensor theory to canonical Einstein gravity. The result
holds under the assumptions that the conformal factor $\Omega \left( \phi 
\right) $ is a monotonic function of the scalar field $\phi$, and that a
global foliation of the spacetime manifold with spacelike hypersurfaces
exists, but it does not require asymptotic flatness. Conformal invariance
of the quasi--local mass was previously found in another generalization to
scalar--tensor theories of the quasi--local mass (Chan,
Creighton and Mann 1996).

The conformal transformation technique has been used to derive new
solutions to
scalar--tensor theories from known solutions of Einstein's theory (Van den
Bergh 1980, 1982, 1983{\em a,b,c,d}; Barrow and Maeda 1990).\\ \\ {\bf
Nonlinear gravitational theories:} a yet more general class of theories
than the previous one is described by the Lagrangian density
\be  \label{f}
{\cal L}=f(R) \sqrt{-g} +{\cal L}_{matter} \; ,
\ee
which generates the field equations
\be
\left( \frac{df}{dR} \right) R_{\mu\nu} -\frac{1}{2}\, f(R) g_{\mu\nu}
-\nabla_{\mu}
\nabla_{\nu} \left( \frac{df}{dR} \right) +g_{\mu\nu} \Box \left(  \frac{df}{dR}
\right) = T_{\mu\nu} \; , 
\ee
\be
T_{\mu\nu}=\frac{2}{\sqrt{-g}} \frac{\delta {\cal L}_{matter}}{\delta
g^{\mu\nu}} \; .
\ee
It is claimed in (Magnano and Sokolowski 1994) that the Einstein frame is
the only physical one for this class of theories, using the energy
argument of Sec.~3. The idea
underlying the proof is to expand the function $f(R)$ as
\be
f(R)=R+aR^2+{\mbox ~...}\; , \;\;\;\;\;\;\;\; a>0\; ,
\ee
and then prove a positive energy theorem in the Einstein frame and the
indefiniteness of the energy sign in the Jordan frame.
The occurrence of singularities in higher order theories of gravity of the
form (\ref{f}) has been studied in
(Barrow and Cotsakis 1988; Miritzis and Cotsakis 1996; Kaloper and Olive
1998), both in the Jordan and in the Einstein frame.\\ \\
{\bf Kaluza--Klein theories:} In
classical Kaluza--Klein theories
(Appelquist, Chodos and Freund 1987; Bailin and Love 1987; Overduin
and Wesson 1997), the scalar field (dilaton) has a geometrical
origin and corresponds to the scale factor of the extra spatial
dimensions.
The extra dimensions manifest themselves as matter (scalar fields) in the
4--dimensional spacetime. In
the simplest version of the theory with a single scalar
field\footnote{See
e.g. (Berezin {\em et al.} 1989; Rainer and Zuhk 1996) for Kaluza--Klein
theories with multiple dilatons.},
one starts with the $(4+d)$--dimensional action of vacuum general relativity
\be
\hat{S}=\frac{1}{16 \pi G}\int d^{(4+d)}x  \left( \hat{R}+
\hat{\Lambda} \right) \sqrt{-\hat{g}} \; ,
\ee
where a caret denotes higher--dimensional quantities,
the $(4+d)$--dimensional metric has the form
\be
\left(\hat{g}_{AB}\right)=\left(
\begin{array}{cc}
\hat{g}_{\mu\nu}  & 0 \\
0 & \hat{\phi}_{ab}
\end{array} \right) \; ,
\ee
and $\hat{\Lambda} $ is the cosmological constant of the
$(4+d)$--dimensional
spacetime manifold. The latter is assumed to have the structure
$ M\otimes K$, where $M$ is 4--dimensional and
$K$ is $d$--dimensional. Here the notations depart from those introduced
at the beginning of this paper: the indices $A,
B$,~...~$=0,1,2,3,$~...~,$(4+d)$; 
$\mu,\nu$,~...~$=0,1,2,3$, and $a,b,... =4,5,$~...,~$(4+d)$.
Dimensional reduction and the conformal transformation (\ref{CT}) with
$\Omega =\sqrt{ \phi}$, $\phi =\left| \hat{\phi}_{ab} \right| $, 
together with the redefinition of the scalar field
\be         \label{sigma}
d \sigma= \frac{1}{2} \left( \frac{d+2}{16\pi G\, d} \right)^{1/2} \frac{d
\phi}{\phi}  \; ,
\ee
leads to the Einstein frame action
\be
S=\int d^4 x \left[\frac{R}{16\pi G} -\frac{1}{2}\, \nabla_{\mu}\sigma
\nabla^{\mu} \sigma -V( \sigma ) \right] \sqrt{-g} \; ,
\ee

\be        \label{VKK}
V( \sigma )=\frac{R_K}{16\pi G} \exp \left( -\sqrt{\frac{16\pi G (d+2)}{d}}
\,\sigma \right) +\frac{\hat{\Lambda}}{16\pi G}
\exp \left(-\sqrt{\frac{16\pi Gd}{d+2} }\, \sigma \right) \; ,
\ee
were $R_K$ is the Ricci curvature of the metric on the submanifold $K$.
Note that $\phi$ is dimensionless. However the redefined scalar field
$\sigma $ has the dimensions $\left[ \sigma \right] =\left[ G^{-1/2}
\right] $, and is usually measured in Planck masses.

Unfortunately, the omission of a factor $1/\sqrt{16\pi G}$ in the right
hand side of Eq.~(\ref{sigma}) seems to be common in the
literature on Kaluza--Klein cosmology (cf. (Faraoni, Cooperstock and
Overduin 1995)
and footnote~11 of (Kolb, Salopek and Turner 1990)) and it leads to
a non--canonical kinetic term
$(16 \pi  G )^{-1} \nabla_{\mu}\sigma \nabla^{\mu} \sigma $ instead of
$ \nabla_{\mu}\sigma \nabla^{\mu} \sigma /2 $ in the final
action, and to a dimensionless field $\sigma$ instead of one with the
correct
dimensions $\left[ \sigma \right]=\left[ G^{-1/2} \right]$.
The error is perhaps due to the different notations used
by particle physicists and by relativists; however insignificant it may
appear to be, it profoundly affects the viability of the Kaluza--Klein
cosmological model
considered, since the spectral index of density perturbations is affected
through the arguments of the exponentials in the scalar field potential
(\ref{VKK}) (Faraoni, Cooperstock and Overduin 1995). In the Jordan
frame, the
scalar field originating from the presence of the extra dimensions has
kinetic energy that is negative definite or indefinite and an energy
spectrum
which is unbounded from below, implying that the ground state is unstable
(Maeda 1986{\em a}; Maeda and Pang 1986; Sokolowski and Carr 1986;
Sokolowski
and Golda 1987). These defects are removed by the conformal rescaling
(\ref{CT}) of the 4--dimensional metric. The requirement that
the conformally rescaled system in 4 dimensions has positive definite
energy (a
purely classical argument) singles out a {\em unique} conformal factor.
A proof of the uniqueness in 5--dimensional Kaluza--Klein theory was
given in (Bombelli {\em et al.} 1987) and later generalized to an
arbitrary number of extra spatial dimensions (Sokolowski 1989{\em a,b}).

From a quantum point of view, arguments in favour of the conformal
rescaling have been pointed out (Maeda 1986{\em b}) and, in the context of
10-- and 11--dimensional supergravity, the need for a conformal
transformation in order
to identify the physical fields was recognized
(Scherk and Schwarz 1979; Chamseddine 1981; Dine {\em et al.} 1985). The
requirement that the supersymmetry
transformation of 11--dimensional supergravity take an $SU(8)$ covariant
form leads to the same conformal factor (de Witt and Nicolai 1986). The 
conformal transformation which works as a cure for the dimensionally
reduced  $(4+d)$--dimensional Einstein gravity does not work for the
dimensionally reduced Gauss--Bonnet theory (Sokolowski {\em et al.} 1991).

It is unfortunate that in the literature on Kaluza--Klein theories many
authors
neglected the conformal rescaling and only performed computations in the
Jordan frame. Many results of classical Kaluza--Klein theories should be
reanalysed in the Einstein frame (e.g.
Alvarez and Bel\'{e}n Gavela 1983; Sadhev 1984; Deruelle and Madore 1987;
Van den Bergh and Tavakol 1993; Fabris and Sakellariadou 1997; Kubyshin
and Martin 1995; Fabris and Martin 1993; Chatterjee and Banerjee 1993;
Biesiada 1994).\\ \\
{\bf Torsion gravity:} Theories of gravity with torsion have been studied
in
order to incorporate the quantum mechanical spin of elementary particles,
or in attempts to formulate gauge theories of gravity (Hehl {\em et
al.} 1976). An
example is given by a theory of gravity with torsion, related to string
theories, recently formulated both in
the Jordan and in the Einstein frame
(Hammond 1990, 1996). Torsion acts as a source of the scalar
field; ordinary (i.e. different from the scalar
field appearing in (\ref{specialCT})) matter is added to the theory
formulated in the Jordan or in the
Einstein frame. This possibility differs from a conformal transformation
of the total (gravity plus matter) system to the Einstein frame, and it
does not appear to be legitimate since ordinary matter cannot be created
as an effect of a conformal transformation.
Although
mathematically possible, this procedure appears to be very artificial, and
it has been considered also in (Magnano and Sokolowski 1994) by including a
nonminimal coupling of the scalar field to matter in the Jordan frame. The 
coupling was tuned in such a way that the Einstein frame matter is
minimally coupled to the corresponding scalar field.
The Jordan frame formulation of this theory is unviable because the large
effects of the dilaton contradict the observations (Hammond 1996), and the
Einstein frame version of this theory is the only possible option.

Induced gravity, which is described by the action
\be  \label{inducedgravity}
S=\int d^4x \sqrt{-g} \left[ -\, \frac{\xi}{2} R\phi^2 -\frac{1}{2}
\nabla^{\mu} \phi \nabla_{\mu} \phi - V( \phi) \right] \; ,
\ee
is conformally invariant if $\xi=1/6$. The field equations are
\be
R_{\mu\nu}-\frac{1}{2} g_{\mu\nu} R = -\frac{1}{\xi \phi^2} \left[ \left(
1-4\xi \right) \nabla_{\mu}
\nabla_{\nu} \phi + g_{\mu\nu} \left( 2\xi -\frac{1}{2}  
\right) \nabla^{\alpha} \phi \nabla_{\alpha} \phi - V g_{\mu\nu} +2 \xi
g_{\mu\nu} \phi \Box \phi \right]   \; ,
\ee
\be
\Box \phi -\xi R \phi - \frac{dV}{d\phi} =0 \; .
\ee
Induced gravity with torsion in Riemann--Cartan spacetimes has been
studied in (Park and Yoon 1997), and a generalization of the concept of
conformal invariance has been
formulated.\\ \\
{\bf Superstring theories:} Although superstring theories are not
classical theories of gravity, the effective action in the low
energy limit is used to make predictions in the classical domain, and we
comment upon this. The low--energy effective action for the
bosonic string theory is given by (Callan {\em et al.} 1985)
\be   \label{stringaction}
S=\int d^{10}x \sqrt{-g} \left[ {\mbox e}^{-2\Phi} R+4 \nabla^{\mu}\Phi
\nabla_{\mu} \Phi  \right] + S_{matter} \; ,
\ee
where $\Phi$ is the dimensionless string dilaton and the totally
antisymmetric 3--form
$H_{\mu\nu\lambda}$ appearing in the theory has been set equal to zero
together
with the cosmological constant (however, this is not always the case in
the literature). By means of dimensional reduction and a conformal
transformation,
this model is reduced to 4--dimensional canonical gravity with two scalar
fields:
\be
\psi_1=\frac{1}{\sqrt{16\pi G}} \left( 6\ln b - \frac{\Phi}{2} \right) \; ,
\ee
\be
\psi_2=\sqrt{\frac{3}{8\pi G}} \left( 2\ln b + \frac{\Phi}{2} \right) \; ,
\ee
where $b$ is the radius of the manifold of the compactified extra
dimensions.
The action (\ref{stringaction}) has provided theoreticians with several
cosmological models
(Gasperini, Maharana and Veneziano 1991;
Garcia--Bellido and Quir\`{o}s 1992;
Gasperini and Veneziano 1992;
Gasperini, Ricci and Veneziano 1993;
Gasperini and Ricci 1993;
Copeland, Lahiri and Wands 1994, 1995;
Batakis 1995;
Batakis and Kehagias 1995;
Barrow and Kunze 1997). The issue of which conformal frame is physical in
the low energy limit of string theories was raised in (Dick 1998).

\section{Conformal transformations in cosmology}

\setcounter{equation}{0}
The standard big--bang cosmology based on general relativity is
a very successful description of the universe that we observe,
although cosmological solutions have been studied also in alternative
theories of
gravity. However, the need to solve the horizon, flatness and monopole
problem, and to find a viable mechanism for the generation of density
fluctuations evolving into the structures that we see today (galaxies,
clusters, supeclusters and voids) motivated research beyond the big--bang
model and led to the idea of cosmological inflation (see Linde 1990; Kolb
and Turner 1990; Liddle and Lyth 1993; Liddle 1996 for reviews). There is
no universally accepted
model of inflation, and several scenarios based either on general
relativity or on alternative theories of gravity have been proposed. 
Since many of the alternative theories used involve a conformal
transformation to a new conformal frame, it is natural that the problem of
whether the Jordan or the Einstein frame is the physical one resurfaces in
cosmology,
together with the use of conformal rescalings to simplify the study of the
equations of motion.
It is possible that general relativity behaves as an attractor for
scalar--tensor theories of
gravity, and that a theory which departs from general
relativity at early times in the history of the universe approaches general
relativity during the matter--dominated era
(Garcia--Bellido and Quir\'{o}s 1990;
Damour and Nordvedt 1993{\em a,b};
Mimoso and Wands 1995{\em a}; Oukuiss 1997) or even during inflation
(Bekenstein and Meisels 1978;
Garcia--Bellido and Quir\'{o}s 1990;
Barrow and Maeda 1990;
Steinhardt and Accetta 1990;
Damour and Vilenkin 1996) (unfortunately only the
Jordan frame was considered in (Garcia--Bellido and
Quir\'{o}s 1990; Mimoso and Wands 1995{\em a})). The convergence to
general relativity cannot occur during the radiation--dominated era
(Faraoni 1998).

One of the most important predictions of an inflationary scenario is the
spectral index of density perturbations, which can
already be compared with the observations of cosmic microwave background
anisotropies and of large scale structures (Liddle and Lyth 1993). The
spectral index is, in general,
different in versions of the same scalar--tensor theory formulated
in different conformal frames. For example,
it is known that most classical Kaluza--Klein inflationary models based
on the Jordan frame are allowed by
the observations but are theoretically unviable
(Sokolowski 1989{\em a,b}; Cho 1992) because of the energy argument
discussed in
Sec.~3; on the contrary, their Einstein frame counterparts are
theoretically consistent but they are severely restricted or even
forbidden by the
observations of cosmic microwave background anisotropies
(Faraoni, Cooperstock and Overduin 1995).
In extended (La and Steinhardt 1989; Kolb, Salopek and Turner 1990; 
Laycock and Liddle 1994) and
hyperextended
(Steinhardt and Accetta 1990; Liddle and Wands 1992; Crittenden
and Steinhardt 1992) inflation, differences between the density
perturbations in the two frames have been pointed out
(Kolb, Salopek and Turner 1990).
The existing confusion on the problem of whether the
Jordan or the Einstein frame is the physical one is particularly
evident
in the literature on inflation, and deeply affects the viability of the
inflationary scenarios based on a theory of gravity which has a conformal
transformation as an ingredient. Among these
are extended (La and Steinhardt 1989; Laycock and Liddle 1994)
and hyperextended
(Kolb, Salopek and Turner 1990;
Steinhardt and Accetta 1990;
Liddle and Wands 1992;
Crittenden and Steinhardt 1992) inflation,
Kaluza--Klein (Yoon and Brill 1990; Cho and Yoon 1993; Cho 1994),
$R^2$--inflation (Starobinski
1980; Starobinski 1986; Maeda, Stein--Schabes and Futamase
1989; Liddle and Lyth 1993), soft and induced gravity inflation
(Accetta, Zoller and Turner 1985; Accetta and
Trester 1989; Salopek, Bond and Bardeen 1989).
While several authors completely neglect the problem of which frame is
physical, other authors present calculations
in only one frame, and others again perform calculations in both frames,
without deciding whether one of
the two is physically preferred. Sometimes, the two frames are
implicitely
treated as if they both were simultaneously physical,
and part of the results are presented in the Jordan frame, part in
the Einstein frame. It is often remarked that all models of inflation based on a
first order phase transition can be recast as slow--roll inflation using a
conformal transformation (Kolb, Salopek and
Turner 1990; Kalara, Kaloper and Olive 1990; Turner 1993; Liddle 1996),
but the conformal rescaling is often performed without physical
motivation. The justification for studying the original (i.e. prior to the
conformal
transformation) theory of gravity or inflationary scenario, which often
relies on a specific theory of high energy physics, is then completely
lost in this way. For example, one can start with a perturbatively
renormalizable potential in the Jordan frame and most likely one
ends up with a non--renormalizable potential in the Einstein frame.
The conformal rescaling has even been used to vary the Jordan frame
gravitational theory in order to obtain a pre--determined scalar field
potential in the Einstein frame (Cotsakis and Saich 1994).

It is to be noted that the conformal transformation to a new conformal
frame is
sometimes used as a purely mathematical device to compute cosmological
solutions by reducing the problem to a familiar (and computationally more
convenient) scenario. The conformal transformation technique has been
used to study also cosmological perturbations in generalized
gravity theories
(Hwang 1990; Mukhanov, Feldman and Brandenberger 1992; Hwang 1997{\em a}).
This technique is certainly legitimate and convenient at the classical
level, but it leads to problems when quantum fluctuations of the inflaton
field are computed in the new conformal frame, and the result is mapped
back into the ``old'' frame. Problems arise already at the
semiclassical level (Duff 1981).
This difficulty does not come as a surprise, since
the conformal transformation introduces a mixing
of the degrees of freedom corresponding to the scalar and the
tensor modes. In general, the fluctuations in the two frames are
physically
inequivalent (Fujii and Nishioka 1990; Makino and Sasaki 1991; Nishioka
and Fujii 1992; Fakir and Habib 1993; Fabris and Tossa 1997).
There is ambiguity in the choice of vacuum states for the quantum fields:
if a vacuum is chosen in one frame, it is unclear into what state the
field is mapped in the other conformal frame, and one will end up, in
general, with two
different quantum states. The use of gauge--invariant quantities does not
fix this problem (Fakir, Habib and Unruh 1992).
The problem that plagues quantum fluctuations becomes relevant for
present--day observations because the quantum perturbations eventually
become classical (Kolb and Turner 1990; Liddle and Lyth 1993; Tanaka and
Sakagami 1997) and
seed galaxies, clusters and superclusters.

Although the problem is not solved in general, the situation is not so
bad in certain specific inflationary scenarios. In (Sasaki 1986; Makino
and Sasaki
1991; Fakir, Habib and Unruh 1992),
chaotic inflation with the quartic potential $V=\lambda \phi^4$ and
nonminimal coupling of the scalar field was studied, and it was found that
the amplitude of the density perturbations does not change under the
conformal
transformation. This result, however, relies on the assumption that one can
split the inflaton field into a classical background plus quantum
fluctuations
(preliminary results when the decomposition is not possible have been
obtained
in (Nambu and Sasaki 1990)).
Under slow--roll conditions in induced gravity inflation, the spectral
index of density perturbations is frame--independent to first order in the
slow--roll parameters (Kaiser 1995{\em a}).
When the expansion of the universe is de Sitter--like,
$a(t) \propto $e$^{Ht}$, $\dot{H} \approx 0$, it was found that the
magnitude of the two--point correlation function is affected by the
conformal
transformation, but its dependence on the wavenumber, and consequently
also the spectral index, is not affected (Kaiser 1995{\em b}). The
spectral indices
differ in the two conformal frames when 
the expansion of the scale factor is close to a power
law\footnote{If inflation occurs in the early universe, it
is not necessarily of
the slow--roll type. The most well studied case of inflation without slow
rolling is power law inflation which occurs for exponential potentials,
obtained in almost
all theories formulated in the Einstein frame.} (Kaiser 1995{\em b}); often,
workers in the field have not been sufficiently careful in this
regard. Certain gauge--invariant quantities related to the cosmological
perturbations turn out to be also conformally invariant under a
mathematical condition satisfied by power law inflation and by the
pole--like inflation encountered in the pre--big bang scenario of low
energy string theory (Hwang 1997{\em b}).

At the level of the classical, {\em unperturbed}
cosmological model, the occurrence of slow--roll inflation in the
Einstein frame does
not necessarily imply that inflation occurs also in the Jordan frame, or
that
it is of the slow--roll type, and  the expansion law of the scale factor is
in
general different in the two conformal frames\footnote{For extended
inflation
in Brans--Dicke theory with $\omega >>1$, it has been proved that
slow--roll
inflation in the Einstein frame implies slow--roll inflation in the Jordan
frame (but not viceversa) (Lidsey 1992; Green and Liddle 1996).}
(see Abreu, Crawford and Mimoso 1994 for an example).

Possible approaches to this problem are outlined in
(Fakir and Habib 1993).
Even if the same expansion law is achieved in the Jordan and the
Einstein frame, the corresponding scalar field potentials can be quite
different in the two frames. For example, power--law inflation is achieved
by an exponential potential for a minimally coupled scalar field in the
Einstein frame, and by a polynomial potential for its nonminimally coupled
cousin in the Jordan frame (Abreu, Crawford and Mimoso 1994;
Futamase and Maeda 1989; Faraoni 1996{\em b}).

Another cosmologically relevant aspect of the scalar field appearing in
(\ref{specialCT}) is that it may contribute a significant fraction of the
dark matter in the universe
(Cho 1990; Cho
and Yoon 1993; Delgado 1994; McDonald 1993{\em a,b};
Gasperini and Veneziano 1994; Gasperini 1994; Cho and Keum 1998).
If one accepts the idea that the scalar field appearing in the expression
for
the conformal factor (\ref{specialCT}) is the field driving inflation
(Salopek 1992; Cho 1992, 1994), then the inflationary scenario is
completely determined. In fact, the conformal transformation to the
Einstein frame in cosmology leads to either $a)$ an exponential potential
for the scalar field and to power--law inflation; $b)$ a potential with
more than one exponential term in Kaluza--Klein theories
(Yoon and Brill 1990; Cho 1990; Cho and Yoon 1993), and to a kind of
inflation that interpolates
between power--law and de Sitter inflation (Easther 1994).
It is also to be noted that, if  a cosmological constant is present in a
theory formulated in the Jordan frame, the new version of the theory in the
Einstein frame has no cosmological constant
(Collins, Martin and Squires 1989; Fujii 1998; Maeda 1992)
but, instead, it exhibits an exponential term in the potential for the
``new'' scalar field.

The problem of whether a Noether symmetry is preserved by the conformal
transformation has been analysed in
(de Ritis {\em et al.} 1990;
Demianski {\em et al.} 1991;
Capozziello, de Ritis and Rubano 1993;
Capozziello and de Ritis 1993;
Capozziello, de Ritis and Marino 1997;
Capozziello and de Ritis 1996, 1997{\em b}).
The asymptotic evolution to an isotropic state of anisotropic Bianchi
cosmologies in higher order theories with Lagrangian density of the
form $ {\cal L}=f(R) \sqrt{-g} +{\cal L}_{matter}$ was studied in
(Miritzis and Cotsakis 1996) using the conformal rescaling
as a mathematical
tool. This  study is relevant to the issue of cosmic no--hair theorems in
these gravitational theories. In the Einstein frame, a homogeneous
universe with matter satisfying the strong and dominant energy conditions
and with a
scalar field with a potential $V( \phi)$ locally convex
and with zero minimum, can isotropize
only if it is of Bianchi type I, V or VII. This result holds also in the
Jordan
frame if, in addition, the pressure of matter is positive
(Miritzis and Cotsakis 1996).

\section{Experimental consequences of the Einstein frame reformulation of
gravitational theories}

\setcounter{equation}{0}
In most unified field theories, the conformal factor used
in the conformal transformation
is constructed using a physical field present in the gravitational theory
(like a dilaton or a Brans--Dicke field) and therefore
it is not surprising that it has certain physical effects which are, in
principle, susceptible of experimental verification. The reality of the
interaction with gravitational strength described by the dilaton was
already stressed by Jordan (1949; 1955).
The dilaton field in the Einstein frame couples differently to gravity and
to matter (e.g. Horowitz 1990; Garfinkle, Horowitz and Strominger 1991),
and the
anomalous coupling results in a violation
of the equivalence principle. Consider for example the action
(\ref{nonlin2})
plus a matter term in the Jordan frame: after the rescaling (\ref{CT}),
(\ref{17}), (\ref{22}) has been performed, the scalar $\tilde{\phi}$
is minimally coupled to gravity,
but it couples nonminimally to the other forms of matter  via a
field--dependent exponential factor:
\be
S=\int d^4 x \left\{ \sqrt{-\tilde{g}} \left[
\frac{\tilde{R}}{16 \pi G}  -\frac{1}{2} \, \tilde{\nabla}^{\mu}
\tilde{\phi} \tilde{\nabla}_{\mu} \tilde{\phi} \right] + {\mbox e}^{-\alpha
\sqrt{G}\, \tilde{\phi}} {\cal L}_{matter} \right\} \; .
\ee
This leads to a violation of the equivalence principle which can, in
principle, be tested by free fall experiments
(Taylor and Veneziano 1988;
Brans 1988;
Cvetic 1989;
Ellis {\em et al.} 1989;
Cho and Park 1991;
Cho 1992;
Damour and Esposito--Far\`{e}se 1992;
Cho 1994;
Damour and Polyakov 1994{\em a,b};
Brans 1997).
It is probably this anomalous coupling and the subsequent violation of
the equivalence principle that explain the prejudice of many
theoreticians against the use of the Einstein frame (which is not,
however, a matter of taste, but is motivated by the independent energy
arguments of Sec.~3). However, it is well known that although  the
Brans--Dicke scalar couples universally to all forms of ordinary matter
in the Jordan frame, the {\em strong} equivalence principle is violated
in this frame. This is sometimes understood as the fact that gravity
determines a {\em local} value of the effective gravitational
``constant'' $G=\phi^{-1}$ (e.g. Brans 1997). In any case, the dilaton
dependence of the coupling
constants is to be regarded as an important prediction of string theories
in the low energy limit, and as a new motivation for improving the
present precision of tests of the equivalence principle.

By describing the gravitational interaction between two point masses $m_1$
and
$m_2$ with the force law
\be
F=\frac{G m_1 m_2}{r^2}\left( 1+\lambda {\mbox e}^{-\mu r} \right) \; ,
\ee
where $\lambda $ and $\mu $ are, respectively, the strength and the range
of the fifth force, one obtains constraints on the range of these
parameters. Due to the smallness of the values of $\lambda $ allowed by
the theory, the null results of the
experiments looking for a fifth force still leave room for a theory
formulated in the Einstein frame and with anomalous coupling
(Cho 1992, 1994; Damour and Polyakov 1994{\em a,b}; Cho and Keum 1998).

There are also post--Newtonian effects and departures from general
relativity
in the strong gravity regime (Damour and Esposito--Far\`{e}se 1992), as
well as differences in
the gravitational radiation emitted and absorbed as compared to
general relativity
(Eardley 1975;
Will and Eardley 1977;
Will 1977;
Will and Zaglauer 1989;
Damour and Esposito--Far\`{e}se 1992).

If $\alpha ( \phi )=\partial ( \ln \Omega ) /\partial \phi$, where
$\Omega$ is the conformal factor in (\ref{CT}),  then the post--Newtonian
parameters $\gamma$ and $\beta$ (Will 1993) are given by (Damour and
Esposito--Far\`{e}se 1992; Damour and Nordvedt 1993{\em a,b})
\be
\gamma -1=\left. -\, \frac{2 \alpha^2}{1+\alpha^2} \right|_{\phi_0} \; ,
\ee
\be
\beta=1+ \frac{\alpha^2}{2\left( 1+\alpha^2 \right)^2}\, \left.
\frac{\partial^2 ( \ln \Omega )}{\partial \phi^2} \right|_{\phi_0} \; ,
\ee
where $\phi_0=\phi (t_0 )$ is the value of the scalar field at the present
time
$t_0$, and it is assumed that the Brans--Dicke field only depends on time.
The 1$\sigma$
limits on $\gamma $ from the Shapiro time delay experiment in the
Solar System (Will 1993) are $|\gamma -1 |< 2\cdot 10^{-3}$ (which implies
$\alpha^2 <10^{-3}$) and the combination $\eta \equiv 4\beta -\gamma -3$
is subject to the constraint $|\eta|< 5\cdot 10^{-3}$.
By contrast, in a scalar--tensor theory, one expects
$\alpha \approx 1$. This value of $\alpha$ could have been realistic early
in the history of the universe with scalar--tensor gravity converging to
general relativity at a later time during the matter--dominated epoch
(Damour and
Nordvedt 1993{\em a,b}). Accordingly, the Jordan and the Einstein frame
would coincide
today, the rescaling (\ref{CT}) differing from the identity only before the
end of the matter--dominated era.

\section{Nonminimal coupling of the scalar field}

\setcounter{equation}{0}
The material contained in this section is a summary of the state of the
art on issues that have been only partially explored, results whose
consequences are largely unknown, and problems that are still open. We
try to point out the directions that, at present, appear most
promising for future research. The reader should be aware of the fact
that due to
the nature of such a discussion, the selection of topics presented here
does not exhaust all the aspects involved.

The generalization to a curved spacetime of the flat space Klein--Gordon
equation for a scalar field $\phi$,
\be              \label{KleinGordon}
\Box \phi -\xi R \phi -\frac{dV}{d\phi}=0 \; ,
\ee
includes the possibility of an explicit coupling term $\xi R\phi $
between the field $\phi$ and the Ricci curvature
of spacetime (Callan, Coleman and Jackiw 1970). There are many reasons to
believe
that a nonminimal (i.e. $\xi \neq 0$) coupling term appears:
a nonminimal coupling is generated by quantum corrections even if it is
absent
in the classical action, or it is required in order to
renormalize the theory (Freedman, Muzinich and Weinberg 1974;
Freedman and Weinberg 1974). It has also been argued in quantum
field theory in curved spaces that a nonminimal coupling term is to be
expected whenever the spacetime curvature is large. This leads to what we
will
call the  ``$\xi$--problem'', i.e. the problem
of whether physics uniquely determines the value of $\xi$. The
answer to this question is affirmative in many theories; several
prescriptions for the coupling constant $\xi$ exist and they differ
according
to the theory of the scalar field adopted. In general relativity  and
in all metric theories of gravity in which the scalar field $\phi$ has a
non--gravitational origin, the value of $\xi$ is fixed to the value $1/6$
by
the Einstein equivalence principle
(Chernikov and Tagirov 1968;
Sonego and Faraoni 1993;
Grib and Poberii 1995;
Grib and Rodrigues 1995;
Faraoni 1996{\em b}).
This is in contrast with a previous claim that
nonminimal coupling spoils the equivalence principle (Lightman
{\em et al.} 1975).
However this claim has been shown to be based on flawed arguments;
instead, it is the minimal coupling of the scalar field that leads to
pathological behaviour (Grib and Poberii 1995; Grib and Rodrigues 1995).
It is interesting
that the derivation of the value $\xi=1/6$ is completely
independent of conformal
transformations, the conformal structure of spacetime, the spacetime metric
and the field equations for the metric tensor of the theory. The fact that the
conformal coupling constant $\xi=1/6$ emerges from these considerations is
extremely
unlikely to be a coincidence, but at present there is no
satisfactory understanding of the reason why this happens, apart from the
following naive consideration. No preferred length scale is present in the
flat space massless Klein--Gordon equation and therefore no such scale
must appear in the limit of the corresponding curved space massless
equation when small regions of spacetime are considered, if the Einstein
equivalence principle holds.

In all theories formulated in the Einstein frame, instead, the scalar field
is minimally coupled ($\xi =0$) to the Ricci curvature, as is evident from
the actions (\ref{actionBDEframe}), (\ref{mincoupl}), (\ref{lin}),
(\ref{actionEframe}).

In many quantum theories of the scalar field $\phi$ there is a unique
solution
to the $\xi$--problem, or there is a restricted range of values of $\xi$.
If $\phi$ is  a Goldstone boson in a theory with spontaneous symmetry
breaking,
$\xi=0$ (Voloshin and Dolgov 1982).
If $\phi$ represents a composite particle,
the value of $\xi$ should be fixed by the known dynamics of its
constituents:
for example, for the Nambu--Jona--Lasinio model,
$\xi=1/6$ in the large $N$
approximation (Hill and Salopek 1992). In the $O(N)$--symmetric model with
$V=\alpha \phi^4$, in which the constituents of the $\phi$ boson are
scalars
themselves, $\xi$ depends on the coupling constants of the elementary
scalars (Reuter 1994): if the coupling of the elementary scalars is
$\xi_0=0$, then $
\xi\in \left[ -1,0 \right]$ while, if $\xi_0=1/6 $, then $\xi =0 $.
For Higgs scalar fields in the standard model and canonical gravity, the
allowed
range of values of $\xi$ is $\xi \leq 0$, $\xi \geq 1/6$ (Hosotani 1985).
The back reaction of gravity on the stability of the scalar $\phi$ in the
potential $ V( \phi)=\eta \phi^3 $ leads to $\xi=0$ (Hosotani 1985).
The stability of a nonminimally coupled scalar field with the
self--interaction
potential 
\be
V( \phi)=\alpha \phi + m^2 \phi^2/2 +\beta \phi^3 +\lambda \phi^4 -
\Lambda 
\ee 
was shown to restrict the possible values of $\xi $ and of the other
parameters of this model (Bertolami 1987). Quantum corrections lead to a
typical value of $\xi$ of order $10^{-1}$ (Allen 1983; Ishikawa 1983).
In general, in a quantum theory $\xi$ is renormalized together with the
other coupling constants of the theory and the particles' masses
(Birrell and Davies 1980; Nelson and Panangaden 1982;
Parker and Toms 1985; Hosotani 1985; Reuter 1994);
this makes an unambiguous solution of the
$\xi$--problem more difficult. In the context of cosmological inflation, a
significant simplification occurs due to the fact that inflation is a
classical, rather than quantum, phenomen: the energy scale involved is
well below the Planck
scale. The potential energy density of the inflaton field 50 e--folds
before the end of inflation is subject to the constraint $V_{50} \leq 6
\cdot 10^{-11} m_{Pl}^4$, where $m_{Pl}$ is the Planck mass
(Kolb and Turner 1990; Turner 1993; Liddle and Lyth 1993). Moreover, 
the trajectory of the inflaton is peaked around classical trajectories
(Mazenko, Unruh and Wald 1985;
Evans and McCarthy 1985;
Guth and Pi 1985; Pi 1985;
Mazenko 1985{\em a,b};
Semenoff and Weiss 1985).
Nevertheless, attempts have been made to begin the inflationary epoch in
the context of string theory or quantum cosmology.  A running coupling
constant in inflationary
cosmology was introduced in (Hill and Salopek 1992) and used to
improve the chaotic inflationary scenario in (Futamase and Tanaka 1997).
Asymptotically free theories in an external gravitational field described
by
the Lagrangian density
\be
{\cal L}=\sqrt{-g} \left( aR^2+b\, G_{GB} +c \, C_{\alpha\beta\gamma\delta}
\, C^{\alpha\beta\gamma\delta} +\xi R \phi^2 \right) +{\cal L}_{matter} \;
,
\ee
where $G_{GB}$ is the Gauss--Bonnet invariant, have a coupling constant
 $\xi (t) $ that
depends on time and tends to $1/6$ when $t \rightarrow \infty $
(Buchbinder 1986; Buchbinder, Odintsov and Shapiro 1986). In the
renormalization group
approach to grand unification theories in curved spaces it was found that,
at the one loop level, $ \xi (t) \rightarrow 1/6 $ or
$ \xi (t) \rightarrow  \infty $ exponentially
(Buchbinder and Odintsov 1983, 1985;
Buchbinder, Odintsov and Lichzier 1989;
Odintsov 1991; Muta and Odintsov 1991; Elizalde and Odintsov 1994).
However, this result is not free of controversies (Bonanno 1995; Bonanno
and Zappal\`a 1997). 

Nonminimal couplings of the scalar field have been widely used in
cosmology, and therefore the above prescriptions have important
consequences for the
viability of inflationary scenarios. In fact, the nonminimal coupling
constant $\xi$ becomes an extra parameter of inflation, and it is well
known that it affects the viability of many scenarios
(Abbott 1981; Starobinsky 1981; Yokoyama 1988; Futamase
and Maeda 1989; Futamase, Rothman and Matzner 1989;  Amendola, Litterio
and Occhionero 1990;
Accioly and Pimentel 1990; Barroso {\em et al.}
1992; Garcia--Bellido and Linde 1995; Faraoni 1996{\em b}). The occurrence
of inflation in anisotropic spaces is also affected by the
value of $\xi$ (Starobinsky 1981; Futamase, Rothman and Matzner 1989;
Capozziello and de Ritis 1997{\em b}), which is
relevant for the cosmic no--hair theorems. In many papers on inflation,
the nonminimal coupling was used to improve the inflationary scenario;
however,
the feeling is that, in general, it actually works in the opposite
direction
(Faraoni 1997{\em a}). In some cases it may be possible to
compare the spectral
index of
density perturbations predicted by the inflationary theory with
the available observations of cosmic microwave background anisotropies
in order to determine the value of
$\xi $ (Kaiser 1995{\em a}; Faraoni 1996{\em b}), or to
obtain other observational constraints (Fukuyama {\em et al.} 1996).
In cosmology, for chaotic inflation with the potential $V=\lambda \phi^4$,
a nonminimal coupling to the curvature lessens the fine--tuning on the
self--coupling
parameter $\lambda$ imposed by the cosmic
microwave background anisotropies
(Salopek, Bond and Bardeen 1989;
Fakir and Unruh 1990{\em a,b};
Kolb, Salopek and Turner 1990;
Makino and Sasaki 1991),
$\lambda < 10^{-12}$. A nonminimal coupling term can also enhance the
growth of density perturbations (Maeda 1992; Hirai and Maeda 1994; Hirai
and Maeda 1997).
For scalar fields in a Friedmann universe, the long wavelenghts $\lambda$
do not scale with the usual reshift formula $\lambda /\lambda_0=a(t) /a(
t_0)$,
but exhibit diffractive corrections if $\xi \neq 1/6$
(Hochberg and Kephart 1991).

The value of the coupling constant $\xi$ affects also the success of
the so--called ``geometric reheating'' of the universe after inflation
(Bassett and Liberati 1998), which is achieved via a nonminimal coupling
of the inflaton with the Ricci curvature, instead of the usual coupling
to a second scalar field.

The ``late time mild inflationary'' scenario of the universe predicts very
short periods of exponential expansion of the universe interrupting the
matter era (Fukuyama {\em et al.} 1997). The model is based on a massive
nonminimally coupled scalar field
acting as dark matter. The success of the scenario depends on the value of
$\xi$, and a negative sign of $\xi$ is necessary. However, the mechanism
proposed in (Fukuyama {\em et al.} 1997) to achieve late time mild
inflation
turns out to be physically pathological from the point of view of wave
propagation in curved spaces (Faraoni and Gunzig 1998{\em b}). At present,
it is unclear whether alternative mechanisms can succesfully
implement the idea of late time mild inflation.

The case $\xi \neq 0$ for a scalar field in the Jordan frame of higher
dimensional models has been shown to have desirable properties in
shrinking the extra dimensions (Sunahara, Kasai and Futamase 1990;
Majumdar 1997), and has been used also for the Brans--Dicke field in
generalized theories of gravity (Linde 1994; Laycock and Liddle 1994;
Garcia--Bellido and Linde 1995).  Exact solutions in cosmology have been
obtained by using the conformal transformation (\ref{CT}), (\ref{OM}),
(\ref{redefNMC}) and starting from known solutions in the Einstein frame,
in which the scalar field is minimally coupled (Bekenstein 1974; Froyland
1992;  Accioly, Vaidya and Som 1983; Futamase and Maeda 1989; Abreu,
Crawford and Mimoso 1994).  From what we have already said in the previous
sections, it is clear that, in general relativity with a nonminimally
coupled scalar field, the Einstein and the Jordan frames are physically
inequivalent but neither is physically preferred on the basis of energy
arguments.

Nonminimal couplings of the scalar field in cosmology have been explored
also in contexts different from inflation
(Dolgov 1983; Ford 1987; Suen and Will 1988;
Fujii and Nishioka 1990;
Morikawa 1990;
Hill, Steinhardt and Turner 1990;
Morikawa 1991;
Maeda 1992;
Sudarsky 1992;
Salgado, Sudarsky and Quevedo 1996, 1997;
Faraoni 1997{\em b}) during the
matter--dominated era (in the
radiation--dominated era of a Friedmann--Lemaitre--Robertson--Walker
solution, or in any spacetime with Ricci curvature $R=0$, the
explicit coupling of the
scalar field to the curvature becomes irrelevant). In particular, a
nonself--interacting, massless scalar field nonminimally coupled to the
curvature with negative $\xi $ has been considered as a mechanism to damp
the cosmological constant (Dolgov 1983; Ford 1987; Suen and Will 1988) and
solve the cosmological constant problem.

Another property of the nonminimally coupled scalar field is remarkable:
while a big--bang singularity is present in many inflationary scenarios
employing a minimally coupled scalar field, it appears that a
nonminimally coupled scalar is a form of matter that can circumvent
the null energy condition and avoid the initial singularity (Fakir 1998).

From the mathematical point of view, the action (\ref{nonmincoupl})
is the only action such that the nonminimal coupling of $\phi$ to $R$
involves
only the scalar field but not its derivatives, and the coupling is
characterized by a dimensionless constant (Birrell and Davies 1982). The
Klein--Gordon equation arising from the action ({\ref{nonmincoupl})
is conformally invariant if $\xi=1/6$ and $V( \phi)=0$, or
$V( \phi)=\lambda \phi^4$. Many authors choose
to reason in terms of an effective renormalization of the
gravitational coupling constant 
\be
G_{eff}=\frac{G}{1-8\pi G \xi \phi^2} \; .
\ee
If $\phi=\phi (t) $, as in spatially homogeneous cosmologies or in
homogeneous regions of spacetime, then the effective gravitational
coupling $G_{eff}=G_{eff}(t)$ varies on a cosmological time scale.
The possibility of  a negative $G_{eff}$ at high energies, corresponding to
an antigravity regime in the early universe has also been considered
(Pollock 1982; Novello 1982), also at the semiclassical level
(Gunzig and Nardone 1984).

The solution of the $\xi$--problem is also relevant for the problem of
backscattering of waves of the scalar $\phi$ off the background curvature of
spacetime, and the creation of ``tails'' of radiation. If the
Klein--Gordon wave equation (\ref{KleinGordon}) is conformally invariant,
tails are absent in any conformally flat spacetime, including the
cosmologically relevant case of Friedmann--Lemaitre--Robertson--Walker
metrics (Sonego and Faraoni 1992; Noonan 1995).
Other areas of gravitational physics for which the solution of the
$\xi$--problem is relevant include the collapse of scalar fields
(Frolov 1998), the theory of the structure and
stability of boson stars (Van der Bij and Gleiser 1987; Liddle and Madsen
1992; Jetzer 1992), which is linked to inflation by the hypothesis
that particles associated with the inflaton field may survive as dark
matter in the form of boson stars. The $\xi$--problem is also relevant for
the field of classical and quantum wormholes, in
which negative energy fluxes are eliminated by restricting the allowed
range of values of $\xi$ (Ford 1987; Hiscock 1990; Coule 1992; Bleyer,
Rainer and Zhuk 1994). Also the Ashtekar formulation of
general relativity has been studied in the presence of nonminimally coupled
scalar fields using a conformal transformation; the field
equations in these variables are nonpolynomial, in contrast to the
polynomial case of minimal coupling (Capovilla 1992).

\section{Conclusions}

\setcounter{equation}{0}
Conformal transformations are extensively used in classical theories of
gravity, higher--dimensional theories and cosmology. Sometimes, the
conformal transformation is a purely mathematical tool that allows one to
map
complicated equations of motion into simpler equations, and constitutes an
isomorphism between spaces of solutions of these equations. In this sense,
the
conformal transformation is a powerful solution--generating technique. More
often, the conformal transformation to the Einstein frame is a map from a
nonviable classical theory of gravity formulated in the Jordan frame to a
viable one which, however, is not as well motivated as the starting one
from the
physical perspective. A key role in establishing the viability of the
Einstein frame version of the theory is played by the positivity of the
energy and by the existence and stability of a ground state in the
Einstein frame. It is to be remarked that the energy argument of Sec.~3
selecting the Einstein frame as the physical one
is not applicable to quantum theories;  in fact, the positivity of energy
and the energy conditions do not hold for quantum theories. The weak energy
condition is violated by quantum states (Ford and Roman 1992, 1993, 1995)
and a theory can
be unstable in the semiclassical regime (Witten 1982), or not have a
ground
state (e.g. Liouville's theory (D'Hoker and Jackiw 1982)).

Conformal transformations, nonminimal coupling, and the related aspects are
important also for quantum and string theories (e.g. Stahlofen and Schramm
1989~--~see Fulton, Rorlich and Witten 1962 for an early review) and for
statistical mechanics (Dita and Georgescu 1989). For example, the
conformal degree of freedom of a conformally flat metric has been studied
in (Padmanabhan 1988) in order to get insight into the quantization of
gravity in the particularly simple case when the spacetime
metric is conformally flat:
$g_{\mu\nu}=\Omega^2 \eta_{\mu\nu}$. In the context of quantum gravity,
lower--dimensional theories of gravity have been under scrutiny for several
years: when the spacetime dimension is 2 or 3,
the metric has only the conformal degree of freedom
(Brown, Henneaux and Teitelboim 1986), because the Weyl tensor vanishes
and any two--
or three--dimensional metric is conformally equivalent to the
Minkowski spacetime of corresponding dimensionality (Wald 1984).
The properties of the quantum--corrected Vlasov equation under conformal
transformations have been studied in (Fonarev 1994).
A nonminimal coupling of a quantum scalar field in a curved space can
induce
spontaneous symmetry  breaking without negative squared masses
(Madsen 1988; Moniz, Crawford and Barroso 1990; Grib and Poberii 1995).
However, all these topics are beyond the purpose of the present work,
which is limited to classical theories.

Many works that appeared and still appear in the literature are affected
by confusion about the conformal transformation technique and the issue
of which conformal frame is physical. Hopefully, these papers
will be reanalysed in the near future in the updated
perspective on the issue of  conformal transformations summarized
in this article.
A change in the point of view is particularly urgent in the analysis of
experimental tests of gravitational theories: most of the current
literature refers to the Jordan frame formulation of Brans--Dicke and
scalar--tensor theories, but it is the Einstein frame which has been
established to be the physical one. A revision is also needed in the
applications of gravitational theories to inflation; the predicted
spectrum of density perturbations must be computed in the physical frame.
In fact, only in this case it is meaningful to compare the theoretical
predictions with the data from the high precision satellite experiments
which map the anisotropies in the cosmic microwave background -- those
already ongoing ({\em COBE} (Smoot {\em et al.} 1992; Bennet {\em et
al.} 1996), and those planned for the early 2000s (NASA's {\em MAP} (MAP
1998) and ESA's {\em PLANCK} (PLANCK 1998)), and from the observations of
large scale structures.

\section*{Acknowledgments}

We are grateful to M. Bruni for suggestions leading to improvements in the
manuscript. VF acknowledges also Y.M. Cho, S. Sonego,  and the colleagues
at the Department of Physics and Astronomy, University of Victoria, for
helpful discussions. This work was partially supported by EEC grants
numbers PSS*~0992 and CT1*--CT94--0004 and by OLAM, Fondation pour la
Recherche Fondamentale, Brussels.

\clearpage

\section*{References}

\noindent Abbott, L.F. (1981), {\em Nucl. Phys. B} {\bf 185}, 233.
\\Abbott, L.F. and Deser, S. (1982), {\em Nucl.
Phys. B} {\bf 195}, 76.
\\Abramowicz, M.A., Carter, B. and
Lasota, J.P. (1988), {\em Gen. Rel. Grav.} {\bf 20}, 1173.
\\Abramowicz, M.A., Lanza, A., Miller, J.C. and Sonego, S. (1997{\em a}),
{\em Gen. Rel. Grav.} {\bf 29}, 1585.
\\Abramowicz, M.A., Andersson, N., Bruni, M., Gosh, P.. and Sonego, S.
(1997{\em b}), {\em Class. Quant. Grav.} {\bf 14}, L189.
\\Abreu, J.P., Crawford, P. and Mimoso, J.P. (1994),
{\em Class. Quant. Grav.} {\bf 11}, 1919.
\\Accetta, F.S. and Trester, J.S. (1989), {\em
Phys. Rev. D} {\bf 39}, 2854.
\\Accetta, F.S., Zoller, D.J. and
Turner, M.S. (1985), {\em Phys. Rev. D} {\bf 31}, 3046.
\\Accioly, A.J. and Pimentel, B.M. (1990), {\em Can.
J. Phys.} {\bf 68}, 1183.
\\Accioly, A.J., Vaidya, A.N. and Som, M.M. (1983), {\em Phys.
Rev. D} {\bf 27}, 2282.
\\Accioly, A.J., Wichowski, U.F., Kwok, S.F. and
Pereira da Silva, N.L. (1993), {\em Class. Quant. Grav.} {\bf 10},
L215.
\\Allen, B. (1983), {\em Nucl. Phys. B} {\bf 226}, 282.
\\Alonso, J.S., Barbero, F., Julve, J. and
Tiemblo, A. (1994), {\em Class. Quant. Grav.} {\bf 11}, 865.
\\Alvarez, E. and Bel\'{e}n Gavela, M. (1983), {\em Phys.
Rev. Lett.} {\bf 51}, 931.
\\Amendola, L., Bellisai, D. and
Occhionero, F. (1993), {\em Phys. Rev. D} {\bf 47}, 4267.
\\Amendola, L., Capozziello, S., Occhionero, F. and Litterio, M.
(1992), {\em Phys. Rev. D} {\bf 45}, 417.
\\Amendola, L., Litterio, M. and Occhionero, F. (1990), {\em Int.
J. Mod. Phys. A} {\bf 5}, 3861.
\\Appelquist, T. and Chodos, A. (1983), {\em
Phys. Rev. Lett.} {\bf 50}, 141.
\\Appelquist, T.,  Chodos, A. and
Freund, P.G.O. (Eds.) (1987), {\em Modern Kaluza--Klein
Theories}, Addison--Wesley, Menlo Park.
\\Bailin, D. and Love, A. (1987), {\em Rep. Prog. Phys.}
{\bf 50}, 1087.
\\Barros, A. and Romero, C. (1998), {\em Phys. Lett. A} {\bf 245}, 31.
\\Barroso, A., Casayas, J., Crawford, P., Moniz, P. and
Nunes, A. (1992), {\em Phys. Lett. B} {\bf 275}, 264.
\\Barrow, J.D. (1993), {\em Phys. Rev. D}
{\bf 47}, 5329.
\\Barrow, J.D. and Cotsakis, S. (1988), {\em Phys. Lett. B} {\bf 214},
515.
\\Barrow, J.D. and Kunze, K.E. (1997), preprint hep--th/9710018.
\\Barrow, J.D. and Maeda, K. (1990), {\em Nucl.
Phys. B} {\bf 341}, 294.
\\Barrow, J.D., Mimoso, J.P. and de
Garcia Maia, M.R. (1993), {\em Phys. Rev. D} {\bf 48}, 3630.
\\Bassett, A.B. and Liberati, S. (1998), {\em Phys. Rev. D} {\bf 58},
021302.
\\Batakis, N.A. (1995), {\em Phys. Lett. B} {\bf 353}, 450.
\\Batakis, N.A. and Kehagias, A.A. (1995), {\em Nucl. Phys. B} {\bf 449},
248--264;
\\Bateman, M. (1910), {\em Proc. Lon. Math. Soc.}
{\bf 8}, 223.
\\Bekenstein, J.D. (1974), {\em Ann. Phys. (NY)}
{\bf 82}, 535.
\\Bekenstein, J.D. and Meisels, A. (1978),
{\em Phys. Rev. D} {\bf 18}, 4378.
\\Belinskii, V.A. and Khalatnikov,
I.M. (1973), {\em Sov. Phys. JETP} {\bf 36}, 591.
\\Bennet {\em et al.} (1996), {\em Astrophys. J. (Lett.)} {\bf 464}, L1.
\\Berezin, V.A., Domenech, G., Levinas, M.L., Lousto, C.O.
and Um\'{e}rez, N.D. (1989), {\em Gen. Rel. Grav.} {\bf 21}, 1177.
\\Bergmann, P.G. (1968), {\em Int. J. Theor. Phys.}
{\bf 1}, 25.
\\Berkin, A.L. and Maeda, K. (1991), {\em
Phys. Rev. D} {\bf 44}, 1691.
\\Berkin, A.L., Maeda, K. and
Yokoyama, J. (1990), {\em Phys. Rev. Lett.} {\bf 65}, 141.
\\Bertolami, O. (1987), {\em Phys. Lett. B} {\bf 186}, 161.
\\Bicknell, G. (1974), {\em J. Phys. A} {\bf 7}, 1061.
\\Biesiada, M. (1994), {\em Class. Quant. Grav.} {\bf 11}, 2589.
\\Billyard, A. and Coley, A. (1997), {\em Mod.
Phys. Lett. A} {\bf 12}, 2121.
\\Birrell, ND. and Davies, P.C. (1980), {\em Phys. Rev. D} {\bf 22},
322.
\\Birrell, N.D. and Davies, P.C. (1982), {\em Quantum
Fields in Curved Space}, Cambridge University Press, Cambridge.
\\Bleyer, U., Rainer, M. and Zhuk, A.
(1994), preprint gr--qc/9405011.
\\Bombelli, L., Koul, R.K., Kunstatter, G.,
Lee, J. and Sorkin, R.D. (1987), {\em Nucl. Phys. B} {\bf 289},
735.
\\Bonanno A. (1995), {\em Phys. Rev. D} {\bf 52}, 969.
\\Bonanno, A. and Zappal\`a, D. (1997), {\em Phys. Rev. D} {\bf 55},
6135.
\\Bose, S. and Lohiya, D. (1997), preprint IUCAA 44/97.
\\Brans, C.H. (1988), {\em Class. Quant. Grav.} {\bf 5},
L197.
\\Brans, C.H. (1997), preprint gr-qc/9705069.
\\Brans, C.H. and Dicke, R.H. (1961), {\em Phys. Rev.}
{\bf 124}, 925.
\\Brown, J.D., Henneaux, M. and
Teitelboim, C. (1986), {\em Phys. Rev. D} {\bf 33}, 319.
\\Bruckman, W.F. and Velazquez, E.S.
(1993), {\em Gen. Rel. Grav.} {\bf 25}, 901.
\\Buchbinder, I.L. (1986), {\em Fortschr. Phys.} {\bf 34}, 605.
\\Buchbinder, I.L. and S.D. Odintsov,
S.D. (1983), {\em Sov. J. Nucl. Phys.} {\bf 40}, 848.
\\Buchbinder, I.L. and Odintsov, S.D. (1985),
{\em Lett. Nuovo Cimento} {\bf 42}, 379.
\\Buchbinder, I.L., Odintsov, S.D. and
Lichzier, I.M. (1989), {\em Class. Quant. Grav.}, {\bf 6}, 605.
\\Buchbinder, I.L., Odintsov, S.D. and
Shapiro, I.L. (1986), in {\em Group--Theoretical Methods in Physics},
Markov, M. (Ed.), Moscow. p.~ 115.
\\Buchbinder, I.L., Odintsov, S.D. and Shapiro, I.L. (1992), {\em Effective
Action in Quantum Gravity}, IOP, Bristol.
\\Buchm\"{u}ller, W. and N. Dragon, N. (1989),
{\em Nucl. Phys. B} {\bf 321}, 207.
\\Callan, C.G. Jr., Coleman, S. and
Jackiw, R. (1970), {\em Ann. Phys. (NY)} {\bf 59}, 42.
\\Callan, C.G., Friedan, D., Martinec, E.J. and
Perry, M.J. (1985), {\em Nucl. Phys. B} {\bf 262}, 593.
\\Campbell, B.A., Linde, A.D. and K.
Olive, K. (1991), {\em Nucl. Phys. B} {\bf 355}, 146.
\\Canuto, V., Adams, P.J., Hsieh, S.--H. and Tsiang, E. (1977), {\em Phys.
Rev. D} {\bf 16}, 1643.
\\Capovilla, R. (1992), {\em Phys. Rev. D} {\bf 46}, 1450.
\\Capozziello, S. and de Ritis, R. (1993),
{\em Phys. Lett. A} {\bf 177}, 1.
\\Capozziello, S. and de Ritis, R. (1996), preprint
astro-ph/9605070.
\\Capozziello, S. and de Ritis, R. (1997{\em a}),
{\em Int. J. Mod. Phys. D} {\bf 6}, 491.
\\Capozziello, S. and de Ritis, R. (1997{\em b}), {\em Gen. Rel. Grav.}
{\bf 29}, 1425.
\\Capozziello, S., de Ritis, R. and Marino, A.A. (1997),
{\em Class. Quant. Grav.} {\bf 14}, 3243.
\\Capozziello, S., de Ritis, R. and Rubano, C.
(1993), {\em Phys. Lett. A} {\bf 177}, 8.
\\Capozziello, S., Occhionero, F. and Amendola, L.
(1993), {\em Int. J. Mod. Phys. D} {\bf 1}, 615.
\\Casas, Garcia--Bellido, J. and M. Quir\'{o}s, M.
(1991), {\em Nucl. Phys. B} {\bf 361}, 713.
\\Cecotti, S. (1987), {\em Phys. Lett. B} {\bf 190}, 86.
\\Chamseddine, A.H. (1981), {\em Nucl. Phys. B} {\bf 185}, 403.
\\Chan, K.C.K., Creighton, J.D.E. and Mann, R.B. (1996), {\em Phys. Rev.
D} {\bf 54}, 3892.
\\Chatterjee, S. and Banerjee, A. (1993), {\em Class. Quant. Grav.} {\bf
10}, L1.
\\Chernikov, N.A. and Tagirov, E.A. (1968),
{\em Ann. Inst. H. Poincar\`{e} A} {\bf 9}, 109.
\\Cho, Y.M. (1987), {\em Phys. Lett. B} {\bf 199}, 358.
\\Cho, Y.M. (1990), {\em Phys. Rev. D} {\bf 41}, 2462.
\\Cho, Y.M. (1992), {\em Phys. Rev. Lett.} {\bf 68},
3133.
\\Cho, Y.M. (1994), in {\em Evolution of the
Universe and its Observational Quest}, Yamada, Japan 1993,
Sato, H. (Ed.), Universal Academy Press, Tokyo, p.~99.
\\Cho, Y.M. (1997), {\em Class. Quant. Grav.} {\bf 14}, 2963.
\\Cho, Y.M. and Keum, Y.Y. (1998), {\em Mod. Phys. Lett. A} {\bf 13}, 109.
\\Cho, Y.M. and Park, D.H. (1991), {\em Gen. Rel. Grav.} {\bf
23}, 741.
\\Cho, Y.M. and J.H. Yoon, J.H. (1993), {\em Phys.
Rev. D} {\bf 47}, 3465.
\\Collins, P.D.B., Martin, A.D. and Squires, E.J. (1989), {\em Particle
Physics and Cosmology}, J. Wiley, New York, p.~293.
\\Copeland, E.J., Lahiri, A. and Wands, D. (1994), {\em Phys. Rev. D}
{\bf 50}, 4868.
\\Copeland, E.J., Lahiri, A. and Wands, D. (1995), {\em Phys. Rev. D}
{\bf 51}, 1569.
\\Cotsakis, S. (1993), {\em Phys. Rev. D} {\bf 47},
1437; 
{\em errata} (1994), {\em Phys. Rev. D} {\bf 49}, 1145.
\\Cotsakis, S. (1995), {\em Phys. Rev. D} {\bf 52}, 6199.
\\Cotsakis, S. and Flessas, G. (1993), {\em
Phys. Rev. D} {\bf 48}, 3577.
\\Cotsakis, S. and Saich, P.J. (1994), {\em
Class. Quant. Grav.} {\bf 11}, 383.
\\Coule, D.H. (1992), {\em Class. Quant. Grav.} {\bf 9}, 2352.
\\Crittenden, R. and Steinhardt, P.J.
(1992), {\em Phys. Lett. B} {\bf 293}, 32.
\\Cunningham, E. (1909), {\em Proc. Lon. Math. Soc.}
{\bf 8}, 77.
\\Cvetic, M. (1989), {\em Phys. Lett. B} {\bf 229}, 41.
\\D'Hoker, E. and Jackiw, R. (1982), {\em Phys.
Rev. D} {\bf 26}, 3517.
\\Damour, T. and Esposito--Far\`{e}se, G. (1992), {\em
Class. Quant. Grav.} {\bf 9}, 2093.
\\Damour, T., Gibbons, G. and
Gundlach, C. (1990), {\em Phys. Rev. Lett.} {\bf 64}, 123.
\\Damour, T. and Gundlach, C. (1991), {\em Phys. Rev.
D} {\bf 43}, 3873.
\\Damour, T. and Nordvedt, K. (1993{\em a}), {\em Phys. Rev.
Lett.} {\bf 70}, 2217.
\\Damour, T. and Nordvedt, K. (1993{\em b}), {\em Phys. Rev. D} {\bf 48},
3436.
\\Damour, T. and Polyakov, A.M. (1994{\em a}), {\em Nucl. Phys.
B} {\bf 423}, 532.
\\Damour, T. and Polyakov, A.M. (1994{\em b}), {\em Gen. Rel. Grav.} {\bf
26}, 1171.
\\Damour, T. and Vilenkin, A. (1996), {\em Phys.
Rev. D} {\bf 53}, 2981.
\\del Campo, S. (1992), {\em Phys. Rev. D}
{\bf 45}, 3386.
\\Delgado, V. (1994), preprint ULLFT--1/94, hep--ph/9403247.
\\Demianski, M., de Ritis, R., Marmo, G. Platania, G.,
Rubano, C., Scudellaro, P. and Stornaiolo, P. (1991), {\em Phys. Rev. D}
{\bf 44}, 3136.
\\de Ritis, R., Marmo, G., Platania, G., Rubano, C.,
Scudellaro, P. and Stornaiolo, C. (1990), {\em Phys. Rev. D}
{\bf 42}, 1091.
\\Deruelle, N., Garriga, J. and Verdaguer, E.
(1991), {\em Phys. Rev. D} {\bf 43}, 1032.
\\Deruelle, N. and Madore, J. (1987), {\em Phys. Lett. B} {\bf 186},
25.
\\Deruelle, N. and Spindel, P. (1990), {\em
Class. Quant. Grav.} {\bf 7}, 1599.
\\Deser, S. (1984), {\em Phys. Lett. B} {\bf 134},
419.
\\de Witt, B. and Nicolai, H. (1986), {\em Nucl.
Phys. B} {\bf 274}, 363.
\\Dick, R. (1988), {\em Gen. Rel. Grav.} {\bf 30}, 435.
\\Dicke, R.H.  (1962), {\em Phys. Rev.} {\bf 125},
2163.
\\Dine, M., Rohm, R., Seiberg, N. and Witten, E.
(1985), {\em Phys. Lett. B} {\bf 156}, 55.
\\Dirac, P.A.M. (1973), {\em Proc. R. Soc. Lond. A}
{\bf 333}, 403.
\\Dita, P. and Georgescu, V. (Eds.) (1989), {\em Conformal Invariance and
String Theory}, Proceedings, Poiana Brasov, Romania 1987, Academic Press,
Boston.
\\Dolgov, A.D. (1983), in {\em The Very Early Universe}, Gibbons, G.W.,
Hawking, S.W. and Siklos, S.T.C. (Eds.), Cambridge University Press,
Cambridge.
\\Duff, M.J. (1981), in {\em Quantum Gravity 2: A Second Oxford Symposium},
Isham, C.J., Penrose, R. and Sciama, D.W. (Eds.), Oxford University
Press, Oxford.
\\Eardley, D.M. (1975), {\em Astrophys. J. (Lett.)} {\bf 196}, L59.
\\Easther, R. (1994), preprint NZ--CAN--RE--94/1,
astro-ph/9405034.
\\Elizalde, E. and Odintsov, S.D. (1994), {\em
Phys. Lett. B} {\bf 333}, 331.
\\Ellis, J. {\em et al.} (1989), {\em Phys. Lett. B} {\bf
228}, 264.
\\Epstein, H., Glaser, V. and Jaffe, A.
(1965), {\em Nuovo Cimento} {\bf 36}, 1016.
\\Evans, M. and McCarthy, J.G. (1985), {\em Phys. Rev. D} {\bf 31},
1799.
\\Fabris, J.C. and Martin, J. (1993), {\em Phys. Lett. B} {\bf 316},
476.
\\Fabris, J.C. and Sakellariadou, M. (1997), {\em Class. Quant. Grav.}
{\bf 14}, 725.
\\Fabris, J.C. and Tossa, J. (1997), {\em Gravit. Cosmol.}
{\bf 3}, 165.
\\ Fakir, R. 1998, preprint gr--qc/9810054.
\\Fakir, R. and Habib, S. (1993), {\em Mod. Phys. Lett.
A} {\bf 8}, 2827.
\\Fakir, R., Habib, S. and Unruh, W.G. (1992), {\em
Astrophys. J.} {\bf 394}, 396.
\\Fakir, R. and Unruh, W.G. (1990{\em a}), {\em Phys.
Rev. D} {\bf 41}, 1783.
\\Fakir, R. and Unruh, W.G. (1990{\em b}), {\em Phys.
Rev. D} {\bf 41}, 1792.
\\Faraoni, V. (1996{\em a}), {\em Astrophys. Lett.
Comm.} {\bf 35}, 305.
\\Faraoni, V. (1996{\em b}), {\em Phys. Rev. D} {\bf
53}, 6813.
\\Faraoni, V. (1997{\em a}), in {\em Proceedings of the 7th
Canadian Conference on General Relativity and Relativistic Astrophysics},
Calgary, Canada 1997, Hobill, D. (Ed.), in press.
\\Faraoni, V. (1997{\em b}), {\em Gen. Rel. Grav.} {\bf 29}, 251.
\\Faraoni, V. (1998), preprint IUCAA 22/98, gr--qc/9805057, to appear in {\em
Phys. Lett. A}.
\\Faraoni, V., Cooperstock, F.I. and Overduin, J.M. (1995), {\em Int. J.
Mod. Phys. A} {\bf 4}, 387.
\\Faraoni, V. and Gunzig, E. (1998{\em a}), {\em Astron. Astrophys.} {\bf
332}, 1154.
\\Faraoni, V. and Gunzig, E. (1998{\em b}), preprint IUCAA 23/98.
\\Ferraris, M. (1986), in {\em Atti del 6$^o$ Convegno
Nazionale di Relativit\`a Generale e Fisica della Gravitazione}, Firenze
1984, Modugno, M. (Ed.), Tecnoprint, Bologna, p.~127.
\\Ferraris, M., Francaviglia, M. and Magnano, G.
(1988), {\em Class. Quant. Grav.} {\bf 5}, L95.
\\Ferraris, M., Francaviglia, M. and Magnano, G.
(1990), {\em Class. Quant. Grav.} {\bf 7}, 261.
\\Fierz, M. (1956), {\em Helv. Phys. Acta} {\bf 29}, 128.
\\Fonarev, O.A. (1994), {\em Class. Quant. Grav.}
{\bf 11}. 2597.
\\Ford, L.H. (1987), {\em Phys. Rev. D} {\bf 35}, 2339.
\\Ford, L.H. and Roman, T.A. (1992), {\em Phys.
Rev. D} {\bf 46}, 1328.
\\Ford, L.H. and Roman, T.A. (1993), {\em Phys. Rev. D}
{\bf 48}, 776.
\\Ford, L.H. and Roman, T.A. (1995), {\em Phys. Rev. D}
{\bf 51}, 4277.
\\Freedman, D.Z., Muzinich, I.J. and Weinberg, E.J.
(1974), {\em Ann. Phys. (NY)} {\bf 87}, 95.
\\Freedman, D.Z. and Weinberg, E.J. (1974), {\em Ann. Phys. (NY)}
{\bf 87}, 354.
\\Freund, P.G.O. (1982), {\em Nucl. Phys. B} {\bf
209}, 146.
\\Frolov, A.V. (1998), preprint gr--qc/9806112. 
\\Froyland, J. (1982), {\em Phys. Rev. D} {\bf 25}, 1470.
\\Fujii, Y. (1998), {\em Progr. Theor. Phys.} {\bf 99}, 599.
\\Fujii, Y. and Nishioka, T. (1990), {\em Phys. Rev. D}
{\bf 42}, 361.
\\Fukuyama, T., Hatakeyama, M., Miyoshi, M.,
Morikawa, M. and Nakamichi, A. (1997), {\em Int. J. Mod. Phys. D} {\bf
6}, 69.
\\Fulton, T., Rohrlich, F.  and
Witten, L. (1962{\em a}), {\em Rev. Mod. Phys.} {\bf 34}, 442.
\\Fulton, T., Rohrlich, F.
and Witten, L. (1962{\em b}), {\em Nuovo Cimento} {\bf 26}, 652.
\\Futamase, T. and Maeda, K. (1989), {\em
Phys. Rev. D} {\bf 39}, 399.
\\Futamase, T., Rothman, T. and Matzner, R. (1989), {\em Phys. Rev. D} {\bf
39}, 405.
\\Futamase, T. and Tanaka, M. (1997), preprint
OCHA--PP--95, hep--ph/9704303.
\\Garay, L. and Garcia--Bellido, J. (1993),
{\em Nucl. Phys. B} {\bf 400}, 416.
\\Garcia--Bellido, J. and Linde, A.D. (1995), {\em Phys.
Rev. D} {\bf 51}, 429.
\\Garcia--Bellido, J. and Linde, A.D. (1995),
{\em Phys. Rev. D} {\bf 52}, 6730.
\\Garcia--Bellido, J. and Quir\'{o}s, M.
(1990), {\em Phys. Lett. B} {\bf 243}, 45.
\\Garcia--Bellido, J. and Quir\`{o}s, M.
(1992), {\em Nucl. Phys. B} {\bf 368}, 463.
\\Garcia--Bellido, J. and Wands, D. (1995), {\em
Phys. Rev. D} {\bf 52}, 5636.
\\Garfinkle, D., Horowitz, G. and Strominger, A. (1991), {\em Phys. Rev.
D} {\bf 43}, 3140; {\em erratum} (1992), {\em Phys. Rev. D} {\bf 45},
3888.
\\Gasperini, M. (1994), {\em Phys. Lett. B} {\bf 327},
214.
\\Gasperini, M., Maharana, J. and Veneziano, G. (1991),
{\em Phys. Lett. B} {\bf 272}, 277.
\\Gasperini, M. and Ricci, R. (1993), {\em Class. Quant. Grav.} {\bf 12},
677.
\\Gasperini, M., Ricci, R. and Veneziano, G. (1993), {\em Phys. Lett. B} {\bf
319}, 438.
\\Gasperini, M. and Veneziano, G. (1992), {\em Phys. Lett. B} {\bf 277},
256.
\\Gasperini, M. and Veneziano, G. (1994),
{\em Phys. Rev. D} {\bf 50}, 2519.
\\Geyer, B. and Odintsov, S.D. (1996), {\em Phys. Rev. D} {\bf 53},
7321.
\\Gibbons, G.W. and Maeda, K. (1988), {\em Nucl.
Phys. B} {\bf 298}, 741.
\\Gott, S., Schmidt, H.--J. and
Starobinsky, A.A. (1990), {\em Class. Quant. Grav.} {\bf 7}, 893.
\\Gottl\"{o}ber, S., M\"{u}ller, V. and A.A.
Starobinsky, A.A. (1991), {\em Phys. Rev. D} {\bf 43}, 2510.
\\Green, A.M. and Liddle, A.R. (1996), {\em
Phys. Rev. D} {\bf 54}, 2557.
\\Green, B.,  Schwarz, J.M. and Witten, E.
(1987), {\em Superstring Theory}, Cambridge University Press, Cambridge.
\\Grib, A.A. and Poberii, E.A. (1995), {\em Helv. Phys.
Acta} {\bf 68}, 380.
\\Grib, A.A. and Rodrigues, W.A. (1995), {\em Gravit.
Cosmol.} {\bf 1}, 273.
\\Gross, D.J. and Perry, M.J. (1983), {\em
Nucl. Phys. B} {\bf 226}, 29.
\\Guendelman, E.I. (1992),
{\em Phys. Lett. B} {\bf 279}, 254.
\\Gunzig, E. and Nardone, P. (1984), {\em Phys. Lett. B}
{\bf 134}, 412.
\\Guth, A.H. and Jain, B. (1992), {\em Phys. Rev. D} {\bf 45}, 426.
\\Guth, A.H. and Pi, S.--Y. (1985), {\em Phys. Rev. D} {\bf 32},
1899.
\\Hammond, R.T. (1990), {\em Gen. Rel. Grav.} {\bf 7}, 2107.
\\Hammond, R.T. (1996), {\em Class. Quant. Grav.} {\bf 13}, L73.
\\Harrison, E.R. (1972), {\em Phys. Rev. D} {\bf 6}, 2077.
\\Hawking, S.W. and Horowitz, G.T. (1996), {\em Class. Quant. Grav.} {\bf
13}, 1487.
\\Hehl, E.W., von der Heyde, P., Kerlick, G.D. and
Nester, J.M. (1976), {\em Rev. Mod. Phys.} {\bf 48}, 393.
\\Higgs, P.W. (1959), {\em Nuovo Cimento} {\bf 11}, 816.
\\Hill, C.T. and Salopek, D.S. (1992), {\em Ann.
Phys. (NY)} {\bf 213}, 21.
\\Hill, C.T., Steinhardt, P.J. and Turner, M.S. (1990), {\em Phys. Lett.
B} {\bf 252}, 343.
\\Hirai, T. and Maeda, K. (1993), preprint WU-AP/32/93.
\\Hirai, T. and Maeda, K. (1994), {\em Astrophys. J.} {\bf 431}, 6.
\\Hirai, T. and Maeda, K. (1997), in {\em Proceedings of the 7th Marcel
Grossman
Meeting}, Stanford, USA 1994, World Scientific, Singapore, p.~477.
\\Hiscock, W.A. (1990), {\em Class. Quant. Grav.} {\bf 7},
L35.
\\Hochberg, D. and Kephart, T.W. (1991), {\em Phys.
Rev. Lett.} {\bf 66}, 2553.
\\Hochberg, D. and Kephart, T.W. (1995),
{\em Phys. Rev. D} {\bf 51}, 2687.
\\Holman, R., Kolb, E.W., Vadas, S. and
Wang, Y. (1991), {\em Phys. Rev. D}  {\bf 43}, 995.
\\Holman, R., Kolb, E.W. and Wang, Y.
(1990), {\em Phys. Rev. Lett.} {\bf 65}, 17.
\\Horowitz, G. (1990), in {\em Proceedings of the 12th International
Conference on General Relativity and Gravitation}, Boulder, USA 1989, N.
Ashby, D. Bartlett and W. Wyss eds. (Cambridge University Press,
Cambridge). 
\\Hosotani, Y. (1985), {\em Phys. Rev. D} {\bf 32}, 1949.
\\Hu, Y., Turner, M.S. and
Weinberg, E.J. (1994), {\em Phys. Rev. D} {\bf 49}, 3830.
\\Hwang, J. (1990), {\em Class. Quant. Grav.} {\bf 7}, 1613.
\\Hwang, J. (1996), {\em Phys. Rev. D} {\bf 53}, 762.
\\Hwang, J. (1997{\em a}), {\em Class. Quant. Grav.} {\bf 14},
1981.
\\Hwang, J. (1997{\em b}), {\em Class. Quant. Grav.} {\bf 14},
3327.
\\Iorio, A., O'Raifeartaigh, L., Sachs, I.
and Wiesendanger, C. (1997), {\em Nucl. Phys. B} {\bf 495}, 433.
\\Ishikawa, J. (1983), {\em Phys. Rev. D} {\bf 28},
2445.
\\Jakubiec, A. and Kijowski, J. (1988),
{\em Phys. Rev. D} {\bf 37}, 1406.
\\Jetzer, P. (1992), {\em Phys. Rep.} {\bf 220}, 163.
\\Jordan, P. (1949), {\em Nature} {\bf 164},
637.
\\Jordan, P. (1955), {\em Schwerkraft und Weltall}, F. Vieweg und Sohn,
Braunschweig.
\\Jordan, P. (1959), {\em Z. Phys.} {\bf 157}, 112.
\\Kaiser, D.I. (1995{\em a}), preprint astro--ph/9507048.
\\Kaiser, D.I. (1995{\em b}), {\em Phys. Rev. D} {\bf 52},
4295.
\\Kalara, S., Kaloper, N. and Olive,
K.A. (1990), {\em Nucl. Phys. B} {\bf 341}, 252.
\\Kaloper, N. and K.A. Olive, K.A. (1998), {\em Phys. Rev. D} {\bf 57},
811.
\\Kasper, U. and Schmidt, H.--J. (1989), {\em
Nuovo Cimento B} {\bf 104}, 563.
\\Klimcik, C. (1993), {\em J. Math. Phys.} {\bf 34}, 1914.
\\Klimcik, C.K. and Kolnik, P. (1993), {\em
Phys. Rev. D} {\bf 48}, 616.
\\Kolb, E.W., Salopek, D. and
Turner, M.S. (1990), {\em Phys. Rev. D} {\bf 42}, 3925.
\\Kolb, E.W. and Turner, M.S. (1990), {\em The
Early Universe}, Addison--Wesley, Reading, Mass.
\\Kolitch, S.J. and Eardley, D.M. (1995), {\em Ann.
Phys. (NY)} {\bf 241}, 128.
\\Kubyshin, Yu. and Martin, J. (1995), preprint UB--ECM--PF 95/13, LGCR
95/06/05, DAMPT R95, gr--qc/9507010.
\\Kubyshin, Y., Rubakov, V. and Tkachev, I.
(1989), {\em Int. J. Mod. Phys. A} {\bf 4}, 1409.
\\Kunstatter, G., Lee. H.C. and
Leivo, H.P. (1986), {\em Phys. Rev. D} {\bf 33}, 1018.
\\La, D. and Steinhardt, P.J. (1989), {\em Phys.
Rev. Lett.} {\bf 62}, 376.
\\Lafrance, R. and Myers, R.C. (1995), {\em Phys. Rev. D} {\bf 51}, 2584.
\\Laycock, A.M. and Liddle, A.R. (1994),
{\em Phys. Rev. D}, {\bf 49}, 1827.
\\Levin, J.J. (1995{\em a}), {\em Phys. Rev. D} {\bf 51},
462.
\\Levin, J.J. (1995{\em b}), {\em Phys. Rev. D} {\bf 51}, 1536.
\\Liddle, A.R. (1996), preprint SUSSEX--AST~96/12--1, astro--ph/9612093,
to appear in Proceedings, {\em From Quantum Fluctuations to Cosmological
Structures}, Casablanca, Morocco 1996.
\\Liddle, A.R. and Lyth, D.H. (1993), {\em
Phys. Rep.} {\bf 231}, 1.
\\Liddle, A.R. and Madsen, M.S. (1992), {\em Int. J. Mod. Phys.} {\bf 1},
101.
\\Liddle, A.R. and Wands, D. (1992), {\em
Phys. Rev. D} {\bf 45}, 2665.
\\Lidsey, E.J. (1992), {\em Class. Quant. Grav.}
{\bf 9}, 149.
\\Lightman, A.P. Press, W.H., Price, R.H. and
Teukolsky, S.A. (1975), {\em Problem Book in Relativity and
Gravitation}, Princeton University Press, Princeton NJ, p.~85.
\\Linde, A.D. (1990), {\em Particle Physics and Inflationary Cosmology},
Hardwood, Chur, Switzerland.
\\Linde, A.D. (1994), {\em Phys. Rev. D} {\bf 49}, 748.
\\Lorentz, H.A. (1937), {\em Collected
Papers}, Nijhoff, The Hague, vol. 5, p.~363.
\\Lorentz--Petzold, D. (1984), in {\em Lecture Notes in
Physics}, Vol.~105, C. Hoenselaers, C. and W. Dietz, W. (Eds.), Springer,
Berlin.
\\Lu, H.Q. and Cheng, K.S. (1996), {\em
Astrophys. Sp. Sci} {\bf 235}, 207.
\\Madsen, M.S. (1988), {\em Class. Quant. Grav.} {\bf 5}, 627.
\\Madsen, M.S. (1993), {\em Gen. Rel. Grav.} {\bf 25}, 855.
\\Maeda, K. (1986{\em a}), {\em Class. Quant. Grav.} {\bf 3}, 651.
\\Maeda, K. (1986{\em b}), {\em Phys. Lett. B} {\bf 166}, 59.
\\Maeda, K. (1987), {\em Phys. Lett. B} {\bf 186}, 33.
\\Maeda, K. (1989), {\em Phys. Rev. D} {\bf 39},
3159.
\\Maeda, K. (1992), in {\em Relativistic Astrophysics and
Cosmology}, Proceedings, Potsdam 1991, Gottl\"{o}ber, S., M\"{u}cket,
J.P. and M\"{u}ller, V. (Eds.), World Scientific, Singapore,
p.~157.
\\Maeda, K. and Pang, P.Y.T. (1986), {\em Phys. Lett. B}
{\bf 180}, 29.
\\Maeda,K., Stein--Schabes, J.A. and Futamase, T. (1989), {\em Phys.
Rev. D} {\bf 39}, 2848.
\\Magnano, G. 1995, in {\em Proceedings of the XI Italian
Conference
on General Relativity and Gravitation}, Trieste, Italy 1994, in press
(preprint gr--qc/9511027).
\\Magnano, G., Ferraris, M. and
Francaviglia, M. (1990), {\em J. Math. Phys.} {\bf 31}, 378.
\\Magnano, G. and Sokolowski, L.M. (1994),
{\em Phys. Rev. D} {\bf 50}, 5039.
\\Majumdar, A.S. (1997), {\em Phys. Rev. D} {\bf 55}, 6092.
\\Makino, N. and Sasaki, M. (1991), {\em Progr. Theor.
Phys.} {\bf 86}, 103.
\\MAP homepage (1998) http://map.gsfc.nasa.gov/
\\Mashoon, B. (1993), in {\em Quantum Gravity and Beyond,
Essays in Honour of Louis Witten on His Retirement}, Mansouri, F. and
Scanio, J.  (Eds.), World Scientific, Singapore.
\\Mazenko, G.F. (1985{\em a}), {\em Phys. Rev. Lett.} {\bf 54},
2163.
\\Mazenko, G.F. (1985{\em b}), {\em Phys. Rev. D} {\bf 34}, 2223.
\\Mazenko, G.F., Unruh, W.G. and Wald, R.M. (1985), {\em Phys.
Rev. D} {\bf 31}, 273.
\\McDonald, J. (1993{\em a}), {\em Phys. Rev. D} {\bf 48},
2462.
\\McDonald, J. (1993{\em b}), {\em Phys. Rev. D} {\bf 48},
2573.
\\Mignemi, S. and Schmidt, H.--J. (1995), {\em
Class. Quant. Grav.} {\bf 12}, 849.
\\Mignemi, S. and Whiltshire, D. (1992),
{\em Phys. Rev. D} {\bf 46}, 1475.
\\Mimoso, J.P. and Wands, D. (1995{\em a}), {\em
Phys. Rev. D} {\bf 51}, 477.
\\Mimoso, J.P. and Wands, D (1995{\em b}), {\em Phys.
Rev. D} {\bf 52}, 5612.
\\Miritzis, J.M. and Cotsakis, S. (1996),
{\em Phys. Lett. B} {\bf 383}, 377.
\\Mollerach, S. and Matarrese, S. (1992),
{\em Phys. Rev. D} {\bf 45}, 1961.
\\Moniz, P., Crawford, P. and Barroso, A. (1990),
{\em Class. Quant. Grav.} {\bf 7}, L143.
\\Morikawa, M. (1990), {\em Astrophys. J. (Lett.)} {\bf
362}, L37.
\\Morikawa, M. (1991), {\em Astrophys. J.} {\bf 369}, 20.
\\Mukhanov, V.F., Feldman, H.A. and
Brandenberger, R.H. (1992), {\em Phys. Rep.} {\bf 215}, 203.
\\Muta, T. and Odintsov, S.D. (1991), {\em
Mod. Phys. Lett. A} {\bf 6}, 3641.
\\Mychelkin, E.G. (1991), {\em Astrophys. Sp. Sci.}
{\bf 184}, 235.
\\Nambu, Y. and Sasaki, M. (1990), {\em Progr. Theor. Phys.} {\bf 83},
37.
\\Nelson, B. and Panangaden, P. (1982), {\em
Phys. Rev. D} {\bf 25}, 1019.
\\Nishioka, T. and Fujii, Y. (1992), {\em Phys.
Rev. D} {\bf 45}, 2140.
\\Noonan, T.W. (1995), {\em Class. Quant. Grav.} {\bf
12}, 1087.
\\Nordvedt, K. (1970), {\em Astrophys. J.} {\bf 161}, 1059.
\\Novello, M. (1982), {\em Phys. Lett. A} {\bf 90}, 347.
\\Novello, M. and Elbaz, E. (1994), {\em Nuovo
Cimento} {\bf 109}, 741.
\\Novello, M. and Heintzmann, H. (1984),
{\em Gen. Rel. Grav.} {\bf 16}, 535.
\\Novello, M., Pereira, V.M.C.
and Pinto--Neto, N. (1995), {\em Int. J. Mod. Phys. D} {\bf 4},
673.
\\Novello, M. and J.M. Salim, J.M.  (1979), {\em
Phys. Rev. D} {\bf 20}, 377.
\\Occhionero, F. and Amendola, L. (1994),
{\em Phys. Rev. D} {\bf 50}, 4846.
\\Odintsov, S.D. (1991), {\em Fortschr. Phys.} {\bf 39},
621.
\\Oukuiss, A. (1997), {\em Nucl. Phys. B} {\bf 486}, 413.
\\Overduin, J.M. and Wesson, P.S. (1997), {\em
Phys. Rep.} {\bf 283}, 303.
\\Padmanabhan, T. (1988), in {\em Highlights in
Gravitation and Cosmology}, Proceedings, Goa, India 1987, Iyer, B.R.,
Kembhavi, A.K., Narlikar, J.V. and Vishveshwara, C.V. (Eds.), Cambridge
University Press, Cambridge, p.~156.
\\Page, L. (1936{\em a}), {\em Phys. Rev.} {\bf 49}, 254.
\\Page, L. (1936{\em b}), {\em Phys. Rev.} {\bf 49},  946.
\\Page, L. and Adams, N.I. (1936), {\em Phys. Rev.} {\bf 49}, 466.
\\Park, C.J. and Yoon, Y. (1997), {\em Gen. Rel.
Grav.} {\bf 29}, 765.
\\Parker, L. and Toms, D.J. (1985), {\em Phys. Rev. D}
{\bf 32}, 1409.
\\Pauli, W. (1955), quoted in {\em Schwerkraft und
Weltall}, F. Vieweg und Sohn, Braunschweig.
\\Pauli, W. (1958), {\em Theory of Relativity}, Pergamon
Press, New York, p.~224.
\\Perlick, V. (1990), {\em Class. Quant. Grav.}
{\bf 7}, 1849.
\\Pi, S.--Y. (1985), {\em Nucl. Phys. B} {\bf 252}, 127.
\\Piccinelli, G., Lucchin, F. and Matarrese, S.
(1992), {\em Phys. Lett. B} {\bf 277}, 58.
\\Pimentel, L.O. and Stein--Schabes, J.
(1989), {\em Phys. Lett. B} {\bf 216}, 27.
\\PLANCK homepage (1998) http://astro.estec.esa.nl/SA--general/Projects/Planck
\\Pollock, M.D. (1982), {\em Phys. Lett. B} {\bf 108}, 386.
\\Rainer, M. and Zuhk, A. (1996), {\em Phys. Rev. D}
{\bf 54}, 6186.
\\Reasenberg, R.D. {\em et al.} (1979), {\em
Astrophys. J. (Lett.)} {\bf 234}, L219.
\\Reuter, M. (1994), {\em Phys. Rev. D} {\bf 49}, 6379.
\\Rothman, T. and Anninos, P. (1991),
{\em Phys. Rev. D} {\bf 44}, 3087.
\\Sadhev, D. (1984), {\em Phys. Lett. B} {\bf 137}, 155.
\\Salgado, M., Sudarsky, D. and Quevedo, H. (1996), {\em Phys. Rev. D} {\bf
53}, 6771.
\\Salgado, M., Sudarsky, D. and Quevedo, H. (1997), {\em Phys. Lett. B}
{\bf 408}, 69.
\\Salopek, D.S. (1992), {\em Phys. Rev. Lett.} {\bf 69},
3602.
\\Salopek, D.S., Bond, J.R. and
Bardeen, J.M. (1989), {\em Phys. Rev. D} {\bf 40}, 1753.
\\Sasaki, M. (1986), {\em Progr. Theor. Phys.} {\bf 76}, 1036.
\\Shapiro, L.L. and Takata, H. (1995), {\em Phys. Lett. B} {\bf 361}, 31.
\\Scheel, M.A., Shapiro, S.L. and Teukolsky, S.A. (1995), {\em Phys. Rev. 
D} {\bf 51}, 4236.
\\Scherk, J. and Schwarz, J.H. (1979), {\em Nucl.
Phys. B} {\bf 153}, 61.
\\Schmidt, H.--J. (1987), {\em Astr. Nachr.} {\bf 308}, 183.
\\Schmidt, H.--J. (1988), {\em Phys. Lett. B} {\bf
214}, 519.
\\Schmidt, H.--J. (1990), {\em Class. Quant. Grav.}
{\bf 7}, 1023.
\\Schmidt, H.--J. (1995), {\em Phys. Rev. D} {\bf 52}, 6196.
\\Schneider, P., Ehlers, J. and Falco, E.E. (1992),
{\em Gravitational Lenses}, Springer, Berlin.
\\Semenoff, G. and Weiss, N. (1985), {\em Phys. Rev. D}
{\bf 31}, 699.
\\Shapiro, I.L. and Takata, H. (1995), {\em
Phys. Lett. B} {\bf 361}, 31.
\\Smoot, G.F. {\em et al.} (1992), {\em Astrophys. J. (Lett.)}
{\bf 396}, L1.
\\Sokolowski, L. (1989{\em a}), {\em Class. Quant. Grav.} {\bf 6}, 59.
\\Sokolowski, L. (1989{\em b}), {\em Class. Quant. Grav.} {\bf 6},
2045.
\\Sokolowski, L.M. (1997), in {\em Proceedings of the
14th International Conference on General Relativity and Gravitation},
Firenze, Italy
1995, M. Francaviglia, G. Longhi, L. Lusanna and E. Sorace eds. (World
Scientific, Singapore), P.~337.
\\Sokolowski, L.M. and Carr, B. (1986), {\em Phys. Lett. B} {\bf 176},
334.
\\Sokolowski, L.M. and Golda, Z.A. (1987), {\em Phys. Lett. B} {\bf
195}, 349.
\\Sokolowski, L.M., Golda, Z.A., Litterio, A.M. and Amendola, L. (1991),
{\em Int. J. Mod. Phys. A} {\bf 6}, 4517.
\\Sonego, S. and Faraoni, V. (1992), {\em J.
Math. Phys.} {\bf 33}, 625.
\\Sonego, S. and Faraoni, V. (1993), {\em Class. Quant.
Grav.} {\bf 10}, 1185.
\\Sonego, S. and Massar, M. (1996), {\em Mon. Not.
R. Astr. Soc.} {\bf 281}, 659.
\\Stahlofen, A.A. and Schramm, A.J. (1989),
{\em Phys. Rev. A} {\bf 40}, 1220.
\\Starobinsky, A.A. (1980), {\em Phys. Lett. B} {\bf 91}, 99.
\\Starobinski, A.A. (1981), {\em Sov. Astron. Lett.} {\bf 7}, 36.
\\Starobinsky, A.A. (1986), {\em Sov. Astr. Lett.} {\bf 29}, 34.
\\Starobinsky, A.A. (1987), in {\em Proceedings of the
4th Seminar on Quantum Gravity},  Markov, M.A. and Frolov, V.P.
(Eds.), World Scientific, Singapore.
\\Steinhardt, P.J. and Accetta, F.S.
(1990), {\em Phys. Rev. Lett.} {\bf 64}, 2740.
\\Streater, R. and Wightman, A. (1964), {\em
PCT, Spin and Statistics, and All That}, Benjamin, New York.
\\Sudarsky, D. (1992), {\em Phys. Lett. B} {\bf 281}, 281.
\\Suen, W.--M. and Will, C.M. (1988), {\em Phys. Lett. B} {\bf 205}, 447.
\\Sunahara, K., Kasai, M. and Futamase, T. (1990),
{\em Progr. Theor. Phys.} {\bf 83}, 353.
\\Suzuki, Y. and Yoshimura, M. (1991),
{\em Phys. Rev. D} {\bf 43}, 2549.
\\Synge, J.L. (1955), {\em Relativity: the General Theory}, North Holland,
Amsterdam.
\\Tanaka, T. and Sakagami, M. (1997), preprint
OU--TAP 50, kucp0107, gr--qc/9705054.
\\Tao, Z.--J. and Xue, X. (1992), {\em Phys. Rev. D}
{\bf 45}, 1878.
\\Taylor, T.R. and Veneziano, G. (1988), {\em
Phys. Lett. B} {\bf 213}, 450.
\\Teyssandier, P. (1995), {\em Phys. Rev. D} {\bf
52}, 6195.
\\Teyssandier, P. and Tourrenc, P.
(1983), {\em J. Math. Phys.} {\bf 24}, 2793.
\\Tkacev, I. (1992), {\em Phys. Rev. D}
{\bf 45}, 4367.
\\Tseytlin, A.A. (1993), {\em Phys. Lett. B} {\bf 317}, 559.
\\Turner, M.S. (1993), in {\em Recent Directions in
Particle Theory -- From
Superstrings and Black Holes
to the Standard Model}, Proceedings of the
Theoretical Advanced Study Institute in Elementary
Particle Physics, Boulder,
Colorado 1992, Harvey, J. and Polchinski, J. (Eds.), World Scientific,
Singapore (preprint FERMILAB--Conf--92/313--A, astro--ph/9304012).
\\Turner, M.S. and Weinberg, E.J. (1997), {\em Phys. Rev. D} {\bf 56},
4604.
\\Turner, M.S. and Widrow, L.M. (1988), {\em Phys. Rev. D} {\bf 37}, 2743.
\\Van den Bergh, N. (1980), {\em Gen. Rel. Grav.} {\bf 12}, 863.
\\Van den Bergh, N. (1982), {\em Gen. Rel. Grav.} {\bf 14}, 17.
\\Van den Bergh, N. (1983{\em a}), {\em Gen. Rel. Grav.} {\bf 15}, 441.
\\Van den Bergh, N. (1983{\em b}), {\em Gen. Rel. Grav.} {\bf 15}, 449.
\\Van den Bergh, N. (1983{\em c}), {\em Gen. Rel. Grav.} {\bf 15},
1043.
\\Van den Bergh, N. (1983{\em d}), {\em Gen. Rel. Grav.} {\bf 16}, 2191.
\\Van den Bergh, N. (1986{\em a}), {\em J. Math. Phys.} {\bf 27}, 1076.
\\Van den Bergh, N. (1986{\em b}), {\em Lett. Math. Phys.} {\bf 11},
141.
\\Van den Bergh, N. (1986{\em c}), {\em Gen. Rel. Grav.} {\bf 18}, 649.
\\Van den Bergh, N. (1986{\em d}), {\em Gen. Rel. Grav.} {\bf 18},
1105.
\\Van den Bergh, N. (1986{\em e}), {\em Lett. Math. Phys.} {\bf 12}, 43.
\\Van den Bergh, N. (1988), {\em J. Math. Phys.} {\bf 29}, 1451.
\\Van den Bergh, N. and Tavakol, R.K. (1993), {\em Class. Quant. Grav.}
{\bf 10}, 183.
\\Van der Bij, J.J. and Gleiser, M. (1987), {\em Phys. Lett. B} {\bf 194},
482.
\\Voloshin, M.B. and Dolgov, A.D. (1982), {\em Sov.
J. Nucl. Phys.} {\bf 35}, 120.
\\Wagoner, R.V. (1970), {\em Phys. Rev. D} {\bf 1}, 3209.
\\Wald, R.M. (1984), {\em General Relativity}, Chicago University Press,
Chicago.
\\Wands, D. (1994), {\em Class. Quant. Grav.} {\bf 11}, 269.
\\Weinstein, S. (1996), {\em Phil. Sci.} {\bf 63}, S63.
\\Weyl, H. (1919), {\em Ann. Phys. (Leipzig)} {\bf 59}, 101.
\\Whitt, B. (1984), {\em Phys. Lett. B} {\bf 145}, 176.
\\Will, C.M. (1977), {\em Astrophys. J.} {\bf 214}, 826.
\\Will, C.M. (1993), {\em Theory and Experiment in Gravitational
Physics} (revised edition), Cambridge University Press, Cambridge.
\\Will, C.M. and Eardley, D.M. (1977), {\em Astrophys. J.
(Lett.)} {\bf 212}, L9.
\\Will, C.M. and P.J. Steinhardt, P.J. (1995),
{\em Phys. Rev. D} {\bf 52}, 628.
\\Will, C.M. and Zaglauer, H.W. (1989), {\em Astrophys. J.}
{\bf 346}, 366.
\\Witten, E. (1982), {\em Nucl. Phys. B} {\bf 195}, 481.
\\Wood, R.W., Papini, G. and Cai, Y.Q.
(1989), {\em Nuovo Cimento B} {\bf 104}, 653.
\\Wu, A. (1992), {\em Phys. Rev. D} {\bf 45}, 2653.
\\Xanthopoulos, B.C. and
Dialynas, T.E. (1992), {\em J. Math. Phys.} {\bf 33}, 1463.
\\Yokoyama, J. (1988), {\em Phys. Lett. B} {\bf 212}, 273.
\\Yoon, J.H. and Brill, D.R. (1990), {\em Class.
Quant. Grav.} {\bf 7}, 1253.

\end{document}